\documentclass[aps,prx,twocolumn,superscriptaddress,longbibliography]{revtex4-2}

\usepackage{graphicx}
\usepackage{amsmath}
\usepackage{amssymb}
\usepackage{bm}
\usepackage{xcolor}
\usepackage{hyperref}
\usepackage{comment}

\graphicspath{{Figures/}}

\begin{document}

\title{Momentum-Selective Electron and Spin Dynamics under Ultrafast Photoexcitation}

\author{Gusein~Bedirkhanov}
\email{gusein.bedirkhanov@polytechnique.edu}
\affiliation{CPHT, CNRS, \'Ecole polytechnique, Institut Polytechnique de Paris, 91120 Palaiseau, France}

\author{Nagamalleswararao~Dasari}
\affiliation{Institut f{\"u}r Theoretische Physik, Universit{\"a}t Hamburg, Notkestra{\ss}e   9 , 22607 Hamburg, Germany}

\author{Alexander~I.~Lichtenstein}
\affiliation{Institut f{\"u}r Theoretische Physik, Universit{\"a}t Hamburg, Notkestra{\ss}e   9 , 22607 Hamburg, Germany}
\affiliation{European X-Ray Free-Electron Laser Facility, Holzkoppel 4, 22869 Schenefeld, Germany}
\affiliation{The Hamburg Centre for Ultrafast Imaging, Luruper Chaussee 149, 22761 Hamburg, Germany}

\author{Evgeny~A.~Stepanov}
\affiliation{CPHT, CNRS, \'Ecole polytechnique, Institut Polytechnique de Paris, 91120 Palaiseau, France}
\affiliation{Coll\`ege de France, 11 place Marcelin Berthelot, 75005 Paris, France}

\begin{abstract}
Recent advances in time-resolved spectroscopies have enabled direct access to the momentum-selective nonequilibrium dynamics of correlated quantum materials, revealing a strongly momentum-dependent response of electrons and collective excitations. 
Interpreting these observations requires a real-time theoretical framework that consistently captures the interplay between strong local electronic correlations and nonlocal collective fluctuations, a capability that remains beyond state-of-the-art nonequilibrium approaches. 
Using a recently developed real-time many-body framework, we resolve the momentum-selective ultrafast dynamics of a photoexcited correlated electron system. 
We predict a transient nodal–antinodal anisotropy in electronic heating, providing a microscopic explanation for the momentum-dependent response debated in time-resolved photoemission and Raman experiments, and identify the nonthermal spectral-weight transfer responsible for the transient antinodal in-gap states observed in ultrafast photoemission. 
We further uncover a momentum-selective magnetic response, in which antiferromagnetic fluctuations undergo a strongly nonthermal, quench-like excitation far above the electronic temperature while preserving their correlation length, before relaxing through a momentum-space magnon cascade toward lower-momentum modes. 
Finally, by tracking the real-time local spin susceptibility, we identify a dynamical, experimentally accessible signature of local-moment formation and its photoinduced melting. 
Our results establish a unified microscopic picture of ultrafast electronic and magnetic dynamics, providing a framework for interpreting momentum-resolved pump–probe experiments.
\end{abstract}

\maketitle

\section{\label{sec:intro}Introduction}

Over the past decade, ultrafast time-resolved experiments on strongly correlated electron systems have uncovered a range of striking phenomena, including light-induced superconductivity~\cite{fausti2011}, metastable hidden phases~\cite{stojchevska2014}, the ultrafast melting of magnetic orders~\cite{afanasiev2019} and insulator-metal transition~\cite{Iwai2003}. 
Beyond probing these systems, ultrafast protocols allow matter to be actively manipulated, engineering new states on demand, for instance through Floquet engineering~\cite{oka2019, delatorre2021}. 
These advances open new opportunities for controlling quantum materials and exploiting their functionalities in future ultrafast technologies~\cite{basov2017}.

The development of advanced time-resolved spectroscopic techniques, such as angle-resolved photoemission spectroscopy (ARPES)~\cite{zonno2021, sobota2021}, resonant inelastic X-ray scattering (RIXS)~\cite{dean2016, cao2019, mazzone2021, jost2026}, and Raman scattering~\cite{Katsumi2025THzRaman, Gatuingt2026, Gallais2026}, has enabled direct access to momentum-selective nonequilibrium dynamics of electronic and collective excitations.
In correlated materials, where electronic degrees of freedom are strongly coupled to collective fluctuations, resolving their momentum and time dependence is essential for understanding the microscopic origin of nonequilibrium phenomena.
A prominent example is the nodal--antinodal dichotomy in cuprates~\cite{norman1998, damascelli2003, kanigel2006, Hashimoto2014}, where strong antiferromagnetic (AFM) fluctuations are associated with the emergence of a pseudogap in the antinodal (AN) regions of the Fermi surface (FS), while coherent quasiparticles survive near the nodal (N) points.
Time-resolved experiments have further revealed that this momentum-dependent distribution of spectral weight around the FS leads to markedly different photoexcitation dynamics~\cite{cilento2018photoenhanced, cilento2014, zonno2021}. 
N quasiparticles respond similarly to electrons in a conventional metal, whereas AN excitations exhibit qualitatively different dynamics, reminiscent of those in an insulator, accompanied by a transient transfer of spectral weight~\cite{cilento2018photoenhanced}.

The momentum selectivity can also be found in collective spin excitations. 
Time-resolved RIXS measurements on photoexcited cuprates and Mott insulating systems have revealed a pronounced momentum dependence of the transient magnetic response, manifested as a nonequilibrium redistribution of magnetic spectral weight across the Brillouin zone (BZ)~\cite{mitrano2020trrixs, cao2019, jost2026, mazzone2021, dean2016}. 
In the electron-doped cuprate compound Nd$_{2-x}$Ce$_x$CuO$_4$, photoexcitation enhances the paramagnon population and redistributes its spectral weight in a momentum-dependent manner~\cite{jost2026}. In the iridates, by contrast, the subsequent magnon relaxation is governed by the available magnetic decay channels. In gapless Sr$_2$IrO$_4$, the steep magnon dispersion toward the zone-center Goldstone mode enables an efficient cascade of magnetic excitations~\cite{mazzone2021, dean2016}, whereas in gapped Sr$_3$Ir$_2$O$_7$ the large spin gap suppresses this relaxation pathway, leading to long-lived transient magnons throughout the BZ~\cite{mazzone2021}.

Beyond the dynamics of magnetic excitations, an equally fundamental question concerns their microscopic origin. 
In paramagnetic correlated systems, increasing the Coulomb interaction drives a crossover from the itinerant Slater regime~\cite{slater1951, rohringer2016}, where magnetism originates from particle–hole instabilities of itinerant electrons, to the localized Heisenberg regime~\cite{anderson1959, chao1977degenerate, chao1977kinetic, macdonald1988, spalek2007}, where preformed local magnetic moments (LMMs) exist prior to magnetic ordering~\cite{chatzieleftheriou_local_2024}. 
In the intermediate-interaction regime, LMMs emerge gradually through a crossover rather than a sharp transition and possess only a finite lifetime~\cite{gunnarsson_breakdown_2017, stepanov_quantum_2018, watzenbock_characteristic_2020, mazitov_local_2021, 
chalupa_fingerprints_2021, stepanov_spin_2022,
gaspard_timescale_2022, mazitov_local_2024}.
Out of equilibrium, this distinction becomes even more subtle, as the local moment itself evolves dynamically, raising the fundamental question of what criterion should define its existence at a given time.

Understanding these phenomena represents a major theoretical challenge.
Although equilibrium properties of correlated systems can be described by a wide range of advanced many-body methods, only a few theoretical frameworks are capable of treating momentum-resolved quantum many-body dynamics in real time while simultaneously capturing the interplay between single-particle excitations and collective charge and spin fluctuations, particularly on ultrafast timescales.
The recently developed real-time dual \textit{GW} (\textit{D-GW}) method~\cite{dasari2025} provides such a framework by treating local correlations nonperturbatively within nonequilibrium dynamical mean-field theory (DMFT)\cite{schmidt2002NeqDMFT, aoki2014}, while incorporating momentum-dependent correlations diagrammatically through the coupling of electrons to spatial charge and spin fluctuations. 
This method has already been applied to the thermalization dynamics of a photoexcited extended Hubbard model near the Mott crossover, revealing distinct energy-transfer pathways between single-particle and collective excitations~\cite{dasari2025}. 
It has also been used to study electrical conductivity near a magnetic instability, demonstrating that an accurate description requires vertex corrections arising from strong magnetic fluctuations, which are absent in DMFT~\cite{dasari_nonlocal_2025}.

In this work, we use \textit{D-GW} to address two complementary aspects of the nonequilibrium dynamics of the correlated electronic systems: the momentum-selective response to photoexcitation and the evolution of LMMs. 
We first analyze the momentum-resolved electronic response to a resonant pump pulse by tracking the dynamics of N and AN quasiparticles. 
We find momentum-selective heating across the FS, with AN quasiparticles showing a retarded heating response relative to the N ones, providing theoretical evidence for the nodal--antinodal anisotropy debated in time-resolved ARPES (tr-ARPES) and ultrafast Raman studies~\cite{Parham2017,Gatuingt2026,Gallais2026}. 
By comparing the transient spectra with their equilibrium counterparts at the same effective temperature, we identify a nonthermal redistribution of spectral weight during the pump, including a transient enhancement driven by strong correlations. 
This offers a microscopic account of the AN in-gap states observed in tr-ARPES on optimally doped cuprates~\cite{cilento2018photoenhanced}.

We further investigate the nonequilibrium dynamics of collective spin excitations and identify distinct behavior of the AFM and long-wavelength magnon modes. 
The AFM mode exhibits a quench-like response with a very high effective temperature and a nearly frozen correlation length, in line with the nonthermal magnon behavior reported by time-resolved RIXS and Raman experiments~\cite{jost2026, Gatuingt2026Thesis}.
On the contrary, the long-wavelength fluctuations follow the electronic subsystem only as long as coherent quasiparticles remain available. 
We further show that the nonequilibrium magnetic response is accompanied by a redistribution of magnetic spectral weight from high- to low-momentum excitations through a momentum-space magnon cascade, consistent with the efficient magnon decay observed in gapless antiferromagnets~\cite{mazzone2021}.

Finally, we study the photoinduced evolution of the local magnetic moment and establish a dynamical and experimentally accessible criterion for its existence based on the slow, interaction-dependent component of the real-time local spin susceptibility, inspired by the two-component picture identified in a semiclassical equilibrium setting in Ref.~\cite{sayad_relaxation_2016}.

Taken together, these results establish a unified theoretical framework for the ultrafast dynamics of correlated electron systems across all interaction strength regimes.
By treating nonequilibrium dynamics together with strong local and nonlocal correlations on an equal footing, it resolves the microscopic mechanisms underlying the electronic and collective excitation response to photoexcitation within a single, consistent picture.
This brings a broad range of ultrafast observations, so far accessed piecewise by complementary probes such as tr-ARPES, ultrafast Raman, and time-resolved RIXS, onto common theoretical ground, establishing a basis for interpreting future time-resolved spectroscopies and anticipating new phenomena in ultrafast correlated matter.
\section{\label{sec:methods}Methods}

\subsection{Model and method}
We consider the half-filled single-band Hubbard model on a square lattice as the simplest prototypical correlated electron system that captures the interplay between local Coulomb interactions and the nonlocal charge and spin collective excitations they induce. 
The Hamiltonian reads
\begin{equation}
H = -J \sum_{\langle ij\rangle,\sigma} c^{\dagger}_{i\sigma} c^{\phantom{\dagger}}_{j\sigma}
+ U \sum_i n_{i\uparrow} n_{i\downarrow},
\label{eq:hubbard}
\end{equation}
where $c^{(\dagger)}_{i\sigma}$ annihilates (creates) an electron with spin $\sigma \in \{\uparrow,\downarrow\}$ on a lattice site $i$, $n_{i\sigma} = c^{\dagger}_{i\sigma}c^{\phantom{\dagger}}_{i\sigma}$ is the corresponding occupation number operator, $J$ is the nearest-neighbor hopping amplitude, $\langle ij \rangle$ restricts the sum to nearest-neighbor pairs, and $U$ is the local Coulomb repulsion.
Throughout this work, energies are measured in units of the half-bandwidth ${D=4J=1}$,
corresponding to a natural unit of time $\hbar/D$, where $\hbar$ is the reduced Planck constant. 
We further adopt natural units $\hbar=e=c=a=1$, where $e$ the elementary charge, $c$ the speed of light, and $a$ the lattice constant. 
We consider ${U\in[0.5,1.5]}$, covering several correlation regimes of the half-filled Hubbard model, from a weakly correlated metal through the metallic pseudogap regime and the metal-to-Mott-insulator crossover~\cite{chatzieleftheriou_local_2024, dasari_nonlocal_2025}.

\begin{figure}[t]
\centering
\includegraphics[width=0.95\columnwidth]{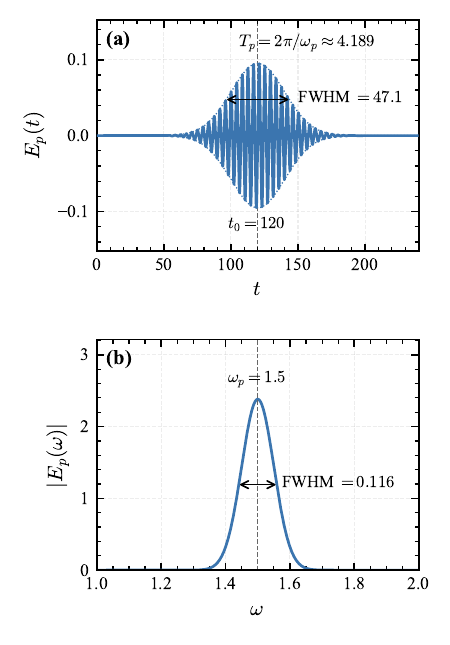}
\caption{Pump pulse used for the interaction strength $U=1.5$. (a) Time-domain electric field $E_p(t)$ with Gaussian envelope. (b) Magnitude of the Fourier spectrum, showing a narrow frequency distribution centered at the resonant carrier frequency $\omega_p = U = 1.5$. The pulse has amplitude $E_0=0.095$ and is centered at $t_0=120$, with a temporal full width at half maximum (FWHM) of $47.1$ and a corresponding spectral FWHM of $0.116$. All quantities are given in units of the half-bandwidth.
\label{fig:pulse}}
\end{figure}

The coupling of electrons to the pump electric field is introduced through the Peierls substitution~\cite{peierls1933}.
For the bare single-particle dispersion of the square lattice,
\begin{equation}
\varepsilon(\mathbf{k}) = -2J(\cos k_x + \cos k_y),
\end{equation}
we adopt the temporal gauge, where the scalar potential is set to zero and the electric field $\mathbf E$ is related to the vector potential as ${\mathbf{A}(t) = -\int_0^t \mathbf{E}_p(\bar{t})\,d\bar{t}}$. 
The vector potential enters as a time-dependent shift of the crystal momentum according to ${\varepsilon(\mathbf{k},t) \equiv \varepsilon\!\left(\mathbf{k}-\mathbf{A}(t)\right)}$. 
We consider a pump pulse linearly polarized along the diagonal of the square lattice, resulting in the time-dependent dispersion
\begin{equation}
    \varepsilon(\mathbf{k},t) = -2J\left[\cos\!\left(k_x-A(t)\right)+\cos \!\left(k_y-A(t)\right)\right].
    \label{eq:dispersion_pulse}
\end{equation}
The pump is characterized by an electric field consisting of a monochromatic carrier modulated by a Gaussian envelope,
\begin{equation}
    E_p(t)=E_0 \exp\left[-\frac{(t-t_0)^2}{2\sigma^2}\right]
    \sin\left[\omega_p(t-t_0)\right],
    \label{eq:pulse}
\end{equation}
where $E_0$, $\omega_p$, $\sigma$, and $t_0$ denote the field amplitude, carrier frequency, temporal width, and pulse center, respectively.
Throughout this work we choose the carrier frequency resonant with the local interaction, ${\omega_p=U}$. 
However, we observed that the obtained results are qualitatively unaffected by moderate variations of the pump parameters, provided that the pulse remains resonant with transitions between the Hubbard bands.
The pump pulse used throughout this work is illustrated in Fig.~\ref{fig:pulse} for the case of ${U=1.5}$.
The pulse spans approximately eleven optical cycles within its temporal full width at half maximum (FWHM) ${\approx 47.1}$, and its Fourier transform exhibits a spectral FWHM ${\approx 0.116}$, corresponding to a relative bandwidth ${\sigma_\omega/\omega_p \approx 3\%}$.
Both the pulse duration and spectral bandwidth are representative of pump pulses employed in tr-ARPES experiments on cuprates, which typically extend over several to a few tens of optical cycles with relative bandwidths of a few percent~\cite{cilento2018photoenhanced, cilento2014}.

We investigate the ultrafast dynamics of the introduced model~\eqref{eq:hubbard} using the real-time \textit{D-GW} method~\cite{dasari2025, dasari_nonlocal_2025}, a nonequilibrium extension of the dual triply irreducible local expansion (D-TRILEX) approach~\cite{Stepanov2019, stepanov2021_1, stepanov2021_2}. 
In \textit{D-GW}, as in other dual diagrammatic extensions of DMFT~\cite{StepanovHDR}, local electronic correlations are treated nonperturbatively within an impurity-like reference problem, which in real time is formulated on the L-shaped Kadanoff--Baym contour~\cite{eckstein2010}. 
The nonlocal correlations are incorporated diagrammatically beyond DMFT, leading to the $GW$-like correction to the impurity self-energy
\begin{align}
\tilde\Sigma^{D\text{-}GW}_{\mathbf k}(t,t')=i
\sum_{\mathbf q, \varsigma}\tilde G_{\mathbf k+\mathbf q}(t,t')\,\tilde W^{\varsigma}_{\mathbf q}(t,t'),
\end{align}
where $\tilde G$ is the nonlocal electronic Green's function and $\tilde W^{\varsigma}$ is the renormalized interaction in the charge (${\varsigma=d}$) and spin (${\varsigma=m\in\{x,y,z\}}$) channel.
This allows the method to self-consistently capture the interplay between strong local correlations and spatial charge and spin fluctuations, which is essential for describing momentum-selective nonequilibrium phenomena. 
As a result, \mbox{\textit{D-GW}} provides direct access to momentum- and frequency-resolved single- and two-particle observables in real time.

\subsection{Time-dependent spectral functions and effective temperatures}
To characterize the nonequilibrium electronic dynamics, we define the lesser ($<$) and greater ($>$) electronic spectral functions through the Wigner transform~\cite{aoki2014} of the corresponding Green's functions,
\begin{align}
    A^{\lessgtr}_{\mathbf k}(\bar t,\omega)
    &=
    \pm\frac{1}{\pi}\,\mathrm{Im}\,G^{\lessgtr}_{\mathbf k}(\bar t,\omega)
    \notag\\&
    =
    \int_0^\infty
    d\tau\,
    e^{i\omega\tau}\,
    G^{\lessgtr}_{\mathbf k}\!\left(\bar t+\tfrac{\tau}{2},\,\bar t-\tfrac{\tau}{2}\right),
    \label{eq:Akw}
\end{align}
with the upper (lower) sign corresponding to the greater (lesser) component.
The retarded spectral function is then given by
\begin{equation}
    A_{\mathbf k}(\bar t,\omega)
    =
    A^>_{\mathbf k}(\bar t,\omega)
    +
    A^<_{\mathbf k}(\bar t,\omega).
\end{equation}

The spectral functions associated with collective charge and spin excitations
are defined analogously from the corresponding charge density ($d$) and magnetic ($m$) susceptibilities. 
In real space these read
\begin{align}
    \chi_{ij}^{d}(t,t')
    &=
    -i\langle T_{\mathcal C}\,
    n_i(t)\,
    n_j(t')
    \rangle,
    \\
    \chi_{ij}^{m}(t,t')
    &=
    -i\langle T_{\mathcal C}\,
    m_i^z(t)\,
    m_j^z(t')
    \rangle,
\end{align}
with
$n_i=n_{i\uparrow}+n_{i\downarrow}$,
$m_i^z=n_{i\uparrow}-n_{i\downarrow}$,
and $T_{\mathcal C}$ the contour-ordering operator. 
The momentum-resolved susceptibilities $\chi_{\mathbf q}^{d/m}(t,t')$ are obtained by spatial Fourier transformation.
Their corresponding lesser and greater spectral functions are
\begin{align}
    A_{\mathbf q}^{d/m,\lessgtr}(\bar t,\omega)
    &=
    -\frac{1}{\pi}\,
    \chi_{\mathbf q}^{d/m,\lessgtr}(\bar t,\omega)
    \notag\\
    &=
    \int_0^\infty
    d\tau\,
    e^{i\omega\tau}\,
    \chi_{\mathbf q}^{d/m,\lessgtr}
    \!\left(\bar t+\tfrac{\tau}{2},
    \bar t-\tfrac{\tau}{2}\right),
    \label{eq:Aqw}
\end{align}
and the retarded spectral functions are
\begin{equation}
    A_{\mathbf q}^{d/m}(\bar t,\omega)
    =
    A_{\mathbf q}^{d/m,>}(\bar t,\omega)
    -
    A_{\mathbf q}^{d/m,<}(\bar t,\omega).
\end{equation}

To track thermalization, we define the nonequilibrium distribution functions
\begin{align}
    F_{\mathbf k}(\bar t,\omega)
    &=
    \ln\!\left[
        \frac{A_{\mathbf k}^{<}(\bar t,\omega)}{A_{\mathbf k}^{>}(\bar t,\omega)}
    \right],
    \\
    F_{\mathbf q}^{d/m}(\bar t,\omega)
    &=
    \ln\!\left[
        \frac{A_{\mathbf q}^{d/m,<}(\bar t,\omega)}{A_{\mathbf q}^{d/m,>}(\bar t,\omega)}
    \right].
\end{align}
In thermal equilibrium, both quantities satisfy the fluctuation--dissipation
relation~\cite{dasari2025,aoki2014},
\begin{equation}
    F(\omega) = -\beta\,(\omega+\mu),
\end{equation}
independent of momentum and of the fermionic or bosonic quantity considered.
Out-of-equilibrium, if $F(\bar t,\omega)$ remains approximately linear over some frequency range,
\begin{equation}
    F(\bar t,\omega) \simeq -\beta_{\mathrm{eff}}(\bar t)\,(\omega + \mu_{\rm eff}(\bar t)),
\end{equation} one can extract the effective inverse temperature $\beta_{\mathrm{eff}}(\bar t)$ and chemical potential $\mu_{\rm eff}(\bar t))$ from a linear fit within that range. Throughout this work, the fit is restricted to the frequency window surrounding the quasiparticle peak at ${\omega=0}$.
The corresponding effective temperature is then defined as
$T_{\mathrm{eff}}(\bar t)=1/\beta_{\mathrm{eff}}(\bar t)$ for both the
electronic and collective degrees of freedom~\cite{dasari2025, dasari2018, murakami2023}.

\subsection{Numerical setup}

We solve the resulting Kadanoff--Baym equations on the L-shaped Keldysh contour using the NESSi library~\cite{schuler2020}. 
The DMFT impurity problem is solved within the real-time non-crossing approximation (NCA)~\cite{eckstein2010}.
The diagrammatic part of the \textit{D-GW} calculation is performed using the implementation presented in Ref.~\cite{dasari2025}.

The pulse and initial-state parameters used for each interaction strength are collected in Table~\ref{tab:params}.
The pump pulse is centered at $t_0=120$ with envelope width $\sigma=20$, and the equations of motion are propagated up to $t_{\max}=240$. All calculations employ real- and imaginary-time steps of $\Delta t=\Delta\tau=0.02$ and a $20\times20$ ${\bf k}$-point mesh spanning the first Brillouin zone, which we found sufficient to converge all dynamical quantities reported in this work.

\begin{table}[h!]
\centering
\caption{Parameters of the four calculations: interaction strength $U$, temperature $T$ of the equilibrium state preceding the pump, and pump field amplitude $E_0$. The carrier frequency is resonant, $\omega_p=U$, and the envelope width is $\sigma=20$ in all cases. All quantities are given in units of the half-bandwidth.}
\label{tab:params}
\begin{ruledtabular}
\begin{tabular}{ccc}
\multicolumn{1}{c}{$U$} &
\multicolumn{1}{c}{$T$} &
\multicolumn{1}{c}{$E_0$} \\
\colrule
0.73 & 0.028 & 0.095 \\
0.85 & 0.036 & 0.125 \\
1.15 & 0.036 & 0.120 \\
1.50 & 0.036 & 0.120 \\
\end{tabular}
\end{ruledtabular}
\end{table}
\section{\label{sec:heating}Results}

\begin{figure*}[t]
\centering
\includegraphics[width=\textwidth]{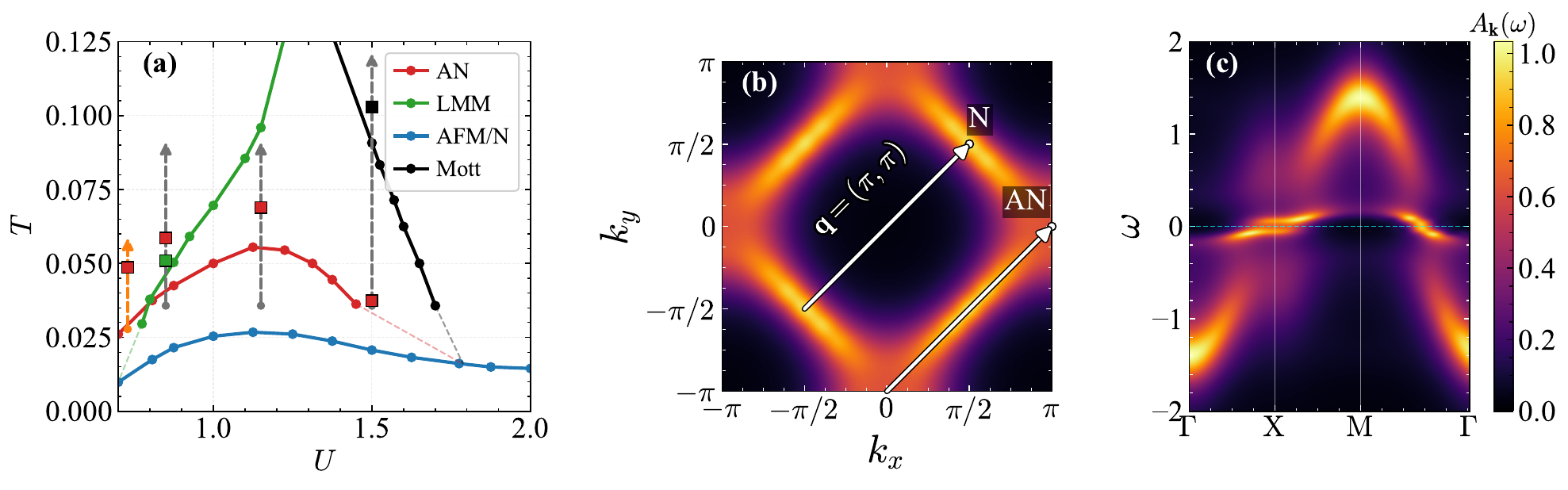}
\caption{Equilibrium properties of the two-dimensional Hubbard model. (a) Equilibrium phase diagram in the $U$--$T$ plane adopted from Ref.~\cite{dasari2025}. The red line denotes the antinodal pseudogap crossover (AN), the blue line the antiferromagnetic (N\'eel) transition (AFM/N), and the black line the Mott transition. Dashed segments indicate extrapolated phase boundaries. The green line (LMM) marks the onset of local magnetic moment formation, determined in this work from the dynamical criterion introduced in Sec.~\ref{sec:llm-criterion}. Vertical dashed arrows show the transient heating trajectories investigated in this work at ${U=0.73}$, $0.85$, $1.15$, and $1.5$. Filled squares along the trajectories indicate characteristic nonequilibrium crossover points extracted from the simulations: red squares denote the transient closure of the antinodal pseudogap, green squares the melting of the local magnetic moment, and the black square the transient crossing of the Mott boundary. (b) Fermi surface in the pseudogap regime at ${U=1.15}$ and ${T=0.04}$. The nodal (${\mathrm{N}=(\pi/2,\pi/2)}$) and antinodal (${\mathrm{AN}=(\pi,0)}$) momenta are indicated and white arrows illustrate the characteristic momentum transfer associated with antiferromagnetic spin fluctuations. (c) Momentum-resolved equilibrium spectral function $A(\mathbf{k},\omega)$ along the high-symmetry path $\Gamma$\,--\,X\,--\,M\,--\,$\Gamma$ for ${U=1.15}$ in the pseudogap regime. The dashed cyan line indicates the Fermi level. Energies and temperatures are expressed in units of the half-bandwidth.}
\label{fig:phase-diagram}
\end{figure*}

\subsection{Nodal--antinodal dichotomy in equilibrium}

Before turning to the nonequilibrium dynamics, we first summarize the key features of the equilibrium electronic structure that provide the reference point for interpreting the photoinduced response. 
Figure~\ref{fig:phase-diagram}\,(a) shows the equilibrium phase diagram in the $U$–$T$ plane obtained within \textit{D-GW}~\cite{dasari_nonlocal_2025}. 
At finite temperatures, increasing the interaction strength $U$ drives the system from a metallic regime toward a Mott insulating state (black line). 
At sufficiently low temperatures, strong AFM correlations emerge across the entire interaction range, signaling the proximity to the N\'eel transition (blue curve). 
The red curve marks the onset of the pseudogap regime induced by AFM fluctuations, where the spectral function develops a minimum at the Fermi level in the antinodal (${\text{AN}=(\pi,0)}$) point of the FS, while the nodal (${\text{N}=(\pi/2,\pi/2)}$) point retains a quasiparticle peak.

The FS in the pseudogap regime, corresponding to the correlated metallic phase at $U=1.15$ and $T=0.04$, is illustrated in Fig.~\ref{fig:phase-diagram}\,(b). 
Strong AFM fluctuations with wave vector ${\mathbf{q}=(\pi,\pi)}$ scatter electrons between nested regions of the FS, leading to a strongly momentum-dependent suppression of the spectral weight. 
This effect is most pronounced in the AN region, where the proximity of the van Hove singularity enhances the effect of AFM scattering and drives the formation of a pseudogap, while the N region remains metallic.

This momentum selectivity is directly reflected in the momentum-resolved equilibrium spectral function shown in Fig.~\ref{fig:phase-diagram}\,(c), computed for the same parameters as the FS in panel~(b).
The nearly flat quasiparticle dispersion in the AN regions, originating from the van Hove singularity, strongly enhances particle-hole scattering between these parts of the FS connected by ${\mathbf{q}=(\pi,\pi)}$, thereby strengthening AFM fluctuations. 
Consequently, coupling of electrons to these AFM fluctuations leads to a splitting of the quasiparticle band at the Fermi level in the AN region. 
In contrast, in the N region, the quasiparticle dispersion remains steep and coherent, resulting in weaker renormalization by spin fluctuations. 
The coexistence of pseudogapped AN states and metallic N quasiparticles gives rise to the characteristic nodal--antinodal dichotomy~\cite{norman1998, damascelli2003, kanigel2006}. 
Upon further lowering the temperature, AFM correlations continue to strengthen until a gap also develops at the N point, driving the system toward an AFM ordered state~\cite{chatzieleftheriou_local_2024}.

\subsection{Photoinduced electron dynamics at N and AN}

\begin{figure*}[t]
\centering
\includegraphics[width=\textwidth]{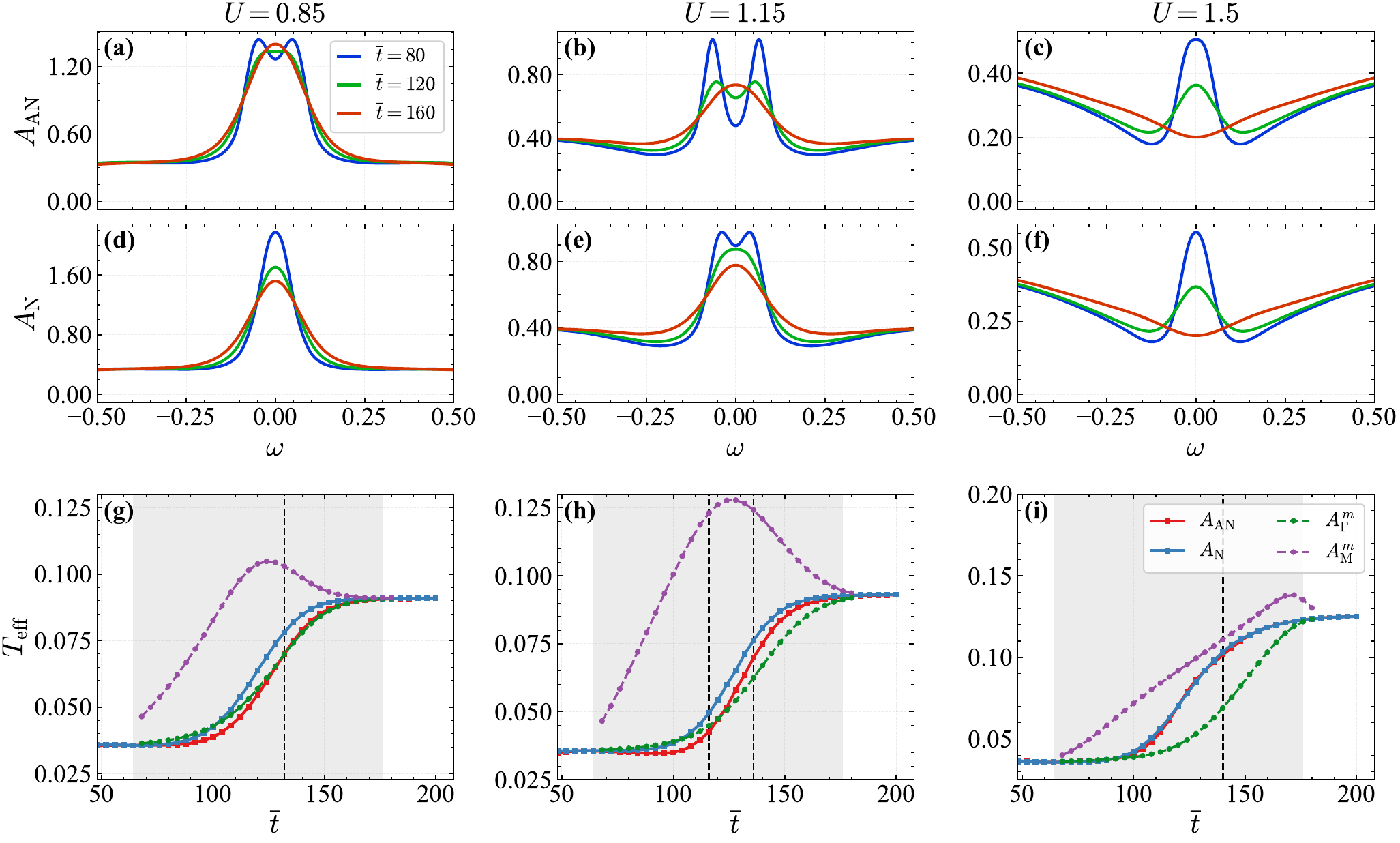}
\caption{Momentum-resolved spectral functions and effective temperatures during the pump excitation, one column per interaction strength. Top row (a)--(c): antinodal spectral function $A_{\mathrm{AN}}(\omega)$ at probe times $\bar t$, evaluated at ${\mathrm{AN}=(\pi,0)}$. Middle row (d)--(f): nodal spectral function $A_{\mathrm{N}}(\omega)$, evaluated at ${\mathrm{N}=(\pi/2,\pi/2)}$. Bottom row (g)--(i): effective temperature $T_{\mathrm{eff}}(\bar t)$ extracted at the antinodal point $A_{\mathrm{AN}}$ (red) and nodal point $A_{\mathrm{N}}$ (blue), together with the magnon effective temperatures $A^{(m)}_{\Gamma}$ (green) and $A^{(m)}_{\mathrm{M}}$ (purple), evaluated at ${\Gamma=(0,0)}$ and ${\mathrm{M}=(\pi,\pi)}$, respectively. For ${U=0.85}$ and $1.15$, the vertical dashed lines mark the gap closing at the antinodal and/or nodal points; for ${U=1.5}$ they mark the Mott transition. The shaded area indicates the effective pulse window. All quantities are given in units of the half-bandwidth.}
\label{fig:pulse-spectral}
\end{figure*}

Having established the equilibrium electronic structure and the origin of the nodal--antinodal dichotomy, we now turn to its nonequilibrium evolution under photoexcitation.
We focus on three representative interaction strengths: the weakly correlated metal (${U=0.85}$), the correlated metal at the center of the pseudogap regime (${U=1.15}$), and the vicinity of the Mott crossover (${U=1.5}$). 
The corresponding photoinduced heating trajectories are indicated by the vertical dashed gray arrows in Fig.~\ref{fig:phase-diagram}\,(a). 
By tracking the momentum-resolved electronic spectra and their effective temperatures throughout the pump pulse, we demonstrate how the equilibrium momentum differentiation governs the momentum-selective redistribution of spectral weight and energy out of equilibrium.

To characterize the photoinduced dynamics, we evaluate the time-dependent spectral function $A_{\mathbf{k}}(\bar{t},\omega)$ at the N and AN momenta, together with the corresponding electronic effective temperature $T_{\mathrm{eff}}(\bar{t})$. 
The results are summarized in Fig.~\ref{fig:pulse-spectral}: the top and middle rows show the evolution of the AN and N spectral functions, respectively, while the bottom row compares the effective temperatures of electrons and magnons.

Before discussing the nonequilibrium dynamics at individual interaction strengths, we first establish how the transient trajectories are represented in the phase diagram. 
As shown in Fig.~\ref{fig:pulse-spectral}, the pump pulse continuously heats the system, such that a single nonequilibrium trajectory at fixed $U$ sweeps through a sequence of effective-temperature regimes, from the low-temperature pseudogapped state to the high-temperature metallic regime (or, for ${U=1.5}$, across the Mott boundary). 
The time-dependent evolution can therefore be mapped onto an effective-temperature trajectory in the phase diagram, allowing the characteristic nonequilibrium crossover points to be identified from a single heating simulation at each interaction strength. 
The corresponding nonequilibrium crossover points are indicated by squares along the heating trajectories in Fig.~\ref{fig:phase-diagram}\,(a).
The red square marks the transient closure of the AN pseudogap, which occurs at an effective temperature significantly higher than the equilibrium pseudogap crossover (red line). 
Similarly, for $U=1.5$, the black square marks the transient crossing of the equilibrium Mott boundary during the heating trajectory. 
These deviations from the equilibrium phase boundaries provide the first indication that the photoinduced dynamics are intrinsically nonthermal and cannot be understood as a simple heating of the equilibrium state.

We first consider the weakly correlated metallic regime at ${U=0.85}$, shown in the first column of Fig.~\ref{fig:pulse-spectral}. 
Before the arrival of the pump, the \textit{D-GW} spectral function exhibits a pronounced nodal--antinodal dichotomy: while the N spectrum displays a sharp quasiparticle peak at the Fermi level (middle panel), the AN spectrum exhibits a pseudogap induced by AFM fluctuations (upper panel). 
As the system is driven out of equilibrium, the AN pseudogap is progressively suppressed, and by the latest probe time coherent quasiparticle peaks have emerged at both momenta, signaling the restoration of a metallic Fermi surface.

The momentum selectivity of electronic heating, shown in Fig.~\ref{fig:pulse-spectral}\,(g), directly reflects the equilibrium electronic structure. 
At the onset of the pump, the metallic N quasiparticles absorb energy considerably more efficiently than the pseudogapped AN states, leading to a rapid increase in the temperature difference between the two regions of the FS. 
This difference remains nearly constant until ${\bar{t}\simeq132}$, when the AN pseudogap collapses. 
Once coherent quasiparticle states are restored at the antinode, the AN heating rate increases, and the temperature difference between the N and AN electrons gradually diminishes.

In the correlated metallic regime at ${U=1.15}$, shown in the second column of Fig.~\ref{fig:pulse-spectral}, the overall nonequilibrium dynamics remains qualitatively similar to that of the weakly correlated metal. 
Before the arrival of the pump, however, both the N and AN spectral functions exhibit pseudogaps. 
During the pulse, the N pseudogap closes first [vertical dashed line at ${\bar{t}\simeq116}$ in Fig.~\ref{fig:pulse-spectral}\,(h)], followed by the AN one [vertical dashed line at ${\bar{t}\simeq136}$]. 
By the end of the pulse, coherent quasiparticle peaks have again developed at both momenta, signaling the restoration of a metallic state.
The momentum-selective heating nevertheless persists. 
Once the nodal pseudogap has collapsed, the N electrons heat more efficiently than their AN counterparts, establishing a temperature difference that remains nearly constant until the antinodal pseudogap also disappears. 
Compared to the weakly correlated metal, however, this transient nodal--antinodal temperature difference is reduced, reflecting the stronger suppression of low-energy spectral weight at both momenta prior to photoexcitation. 
A further distinctive feature is the nonmonotonic evolution of the AN effective temperature, which initially decreases slightly before increasing as the antinodal pseudogap collapses, reflecting the strongly nonthermal evolution of the AN electronic distribution during the early stage of the pump.

The third column of Fig.~\ref{fig:pulse-spectral} shows the results obtained near the Mott crossover, at ${U=1.5}$. 
In this regime, the nonequilibrium electron dynamics changes qualitatively. 
In equilibrium, the vicinity of the Mott crossover is characterized by the formation of well-developed local magnetic moments, which substantially reduces the fraction of itinerant electrons in the system~\cite{chatzieleftheriou_local_2024}. 
As a consequence, the nodal--antinodal dichotomy becomes strongly suppressed: the AN spectrum exhibits only a weak splitting of the quasiparticle peak (top panel), comparable to that at the N point (middle panel). 
During the pump, the coherent quasiparticle peaks initially present at both momenta are progressively suppressed and eventually disappear as the system is driven across the Mott crossover [vertical dashed line at ${\bar{t}\simeq140}$ in Fig.~\ref{fig:pulse-spectral}\,(i)]. 
The N and AN spectra evolve in a nearly identical manner, demonstrating that the momentum-selective dynamics characteristic of the metallic regime are largely washed out once well-developed local magnetic moments dominate the low-energy electronic structure.

Experimentally, the transient electronic temperature in cuprates has so far been extracted predominantly from the N region. 
Tr-ARPES measurements of the photoinduced dynamics, such as those of Ref.~\cite{Parham2017}, provide an estimate of the effective electronic temperature at the node, while extracting a reliable temperature in the AN region remains challenging due to the experimental limitations. 
This limitation has been partially overcome by ultrafast Raman thermometry (e.g., Refs.~\cite{Gatuingt2026, Gallais2026}), which enables selective probing of the response from the N and AN regions of the FS.
Combining the results of these two techniques suggests possible momentum-selective heating, since the N relaxation extracted from tr-ARPES~\cite{Parham2017} is slower than the AN relaxation inferred from Raman measurements~\cite{Gatuingt2026}. 
However, such a comparison remains inconclusive because the two probes couple to different electronic and collective excitations. 
A symmetry-resolved Raman measurement on optimally doped Bi$_2$Sr$_2$CaCu$_2$O$_{8+\delta}$~\cite{Gallais2026} indeed finds a higher quasiparticle temperature in the N channel than in the AN channel, but the magnitude of this difference cannot be unambiguously separated from systematic uncertainties and contributions from magnetic excitations.
Whether the N and AN quasiparticle temperatures exhibit a genuine anisotropy therefore remains an open question. 
Our calculations resolve this ambiguity by providing a microscopic description of the coupled electronic and collective dynamics, demonstrating that the nodal--antinodal temperature anisotropy is an intrinsic property of the photoexcited correlated system, with nodal quasiparticles reaching higher transient temperatures than antinodal excitations.

\begin{figure}[t!]
\centering
\includegraphics[width=\columnwidth]{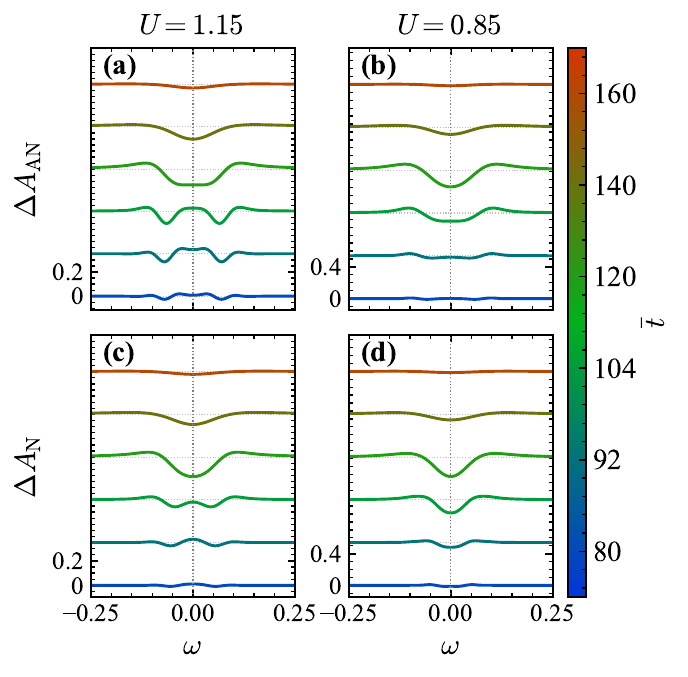}
\caption{Nonthermal contribution to the transient spectral function, ${\Delta A_{\mathbf{k}}(\bar t,\omega) = A_{\mathbf{k}}(\bar t,\omega) - A^{\mathrm{eq}}_{\mathbf{k}}(T_{\mathrm{eff}}(\bar t),\omega)}$, for the weakly correlated metal (${U=1.15}$, left column) and the correlated metal (${U=0.85}$, right column), at the antinodal momentum ${\mathrm{AN}=(\pi,0)}$ (top row, $\Delta A_{\mathrm{AN}}$) and nodal momentum ${\mathrm{N}=(\pi/2,\pi/2)}$ (bottom row, $\Delta A_{\mathrm{N}}$). The comparison is made with equilibrium spectra calculated at the same $T_{\mathrm{eff}}^{\mathrm{loc}}(\bar t)$ extracted from the nonequilibrium local spectral function $A_{\mathrm{loc}}(\bar t,\omega)$. Curves are offset vertically and colored by probe time $\bar t$ (color bar), covering the pulse window. Energies are in units of the half-bandwidth.}
\label{fig:deltaA}
\end{figure}

While the effective-temperature analysis captures the overall energy transfer into the electronic system, it does not distinguish a thermal redistribution from a genuinely nonthermal redistribution of spectral weight. 
To isolate the latter, we compare the transient spectral function $A_{\mathbf{k}}(\bar{t},\omega)$ with the equilibrium spectrum $A^{\mathrm{eq}}_{\mathbf{k}}(T_{\mathrm{eff}}(\bar{t}),\omega)$ evaluated at the same effective temperature $T_{\mathrm{eff}}$ extracted from the local spectral function. 
The resulting difference $\Delta A_{\mathbf{k}}(\bar{t},\omega)=A_{\mathbf{k}}(\bar{t},\omega)-A^{\mathrm{eq}}_{\mathbf{k}}(T_{\mathrm{eff}}(\bar{t}),\omega)$
for the N and AN points at weak coupling ($U=0.85$, left column) and in the correlated metallic regime ($U=1.15$, right column) is shown in Fig.~\ref{fig:deltaA}.

For $U=1.15$, the early-time response exhibits a characteristic two-component structure at both momenta. 
A positive contribution develops around the Fermi level (${\omega=0}$), indicating an excess of quasiparticle spectral weight inside the equilibrium pseudogap, while negative side lobes emerge at the pseudogap edges (${\omega\simeq\pm0.08}$), corresponding to a depletion of spectral weight at these energies. 
The in-gap contribution is substantially stronger at the AN point, where the equilibrium pseudogap is deepest, demonstrating that the strength of the nonthermal response follows the momentum dependence of electronic correlations. 
As the pump proceeds and the pseudogap collapses, the negative redistribution near the gap edges gradually dominates, while the in-gap contribution is suppressed. 
By the end of the pulse, $\Delta A_{\mathbf{k}}$ becomes small over the entire energy range, indicating that the transient spectrum at the later times is well described by an equilibrium spectrum at the corresponding effective temperature.

In contrast, the weakly correlated metal ($U=0.85$) exhibits only a single-component response: the positive in-gap contribution is absent at both the N and AN points, and $\Delta A_{\mathbf{k}}$ consists primarily of a negative redistribution of spectral weight around the Fermi level. 
The absence of the in-gap component reflects the lack of well-developed Hubbard bands, leaving little incoherent Hubbard-band spectral weight available for transfer into low-energy quasiparticle states.
Thus, the transient enhancement of in-gap spectral weight emerges as a distinct signature of strong electronic correlations. 
By contrast, the negative redistribution around the Fermi level appears for all momenta and interaction strengths considered here, indicating a generic nonequilibrium response associated with photoexcitation.

The momentum-dependent transient dynamics revealed by our calculations have a qualitative correspondence with time-resolved spectroscopic observations in optimally doped cuprates~\cite{cilento2018photoenhanced, cilento2014, zonno2021}. 
These experiments show that the N electronic distribution exhibits conventional metallic heating, whereas photoexcitation of the AN region induces transient in-gap quasiparticle states, a hallmark of photoinduced Mottness, namely the transient release of correlation-frozen AN spectral weight. 
In particular, Ref.\cite{cilento2018photoenhanced} demonstrated that the antinodal response cannot be explained by simple gap filling, gap closing, or purely thermal broadening of the electronic distribution, but instead requires a genuine transient enhancement of spectral weight at the Fermi level. 
The nonthermal redistribution observed in Fig.~\ref{fig:deltaA} reproduces this experimentally observed enhancement of in-gap spectral weight and, additionally, resolves the accompanying depletion at the pseudogap edges, revealing the microscopic spectral-weight transfer from the Hubbard bands into low-energy quasiparticle states.

\subsection{Photoinduced spin dynamics}

At the temperatures considered, collective charge excitations remain negligible compared with their magnetic counterparts, and we therefore focus exclusively on the spin channel. 
To characterize the nonequilibrium magnetic response, we extract effective temperature of spin excitations at two representative momenta. 
The long-wavelength mode at ${\Gamma=(0,0)}$ corresponds to a low-energy particle–hole excitation, whereas the mode at ${\text{M}=(\pi,\pi)}$ represents antiferromagnetic fluctuations. 
The resulting effective temperatures are shown in the bottom row of Fig.~\ref{fig:pulse-spectral}. 
Since the spin susceptibility decays substantially more slowly than the electronic Green's function, extracting reliable effective temperatures requires a sufficiently long relative-time window, restricting the analysis to a limited interval around the pump.
The corresponding data are therefore omitted at early and late times.

We find that the transient magnetic response is dominated by the AFM mode, whose effective temperature shows the largest deviations from the electronic one and exceeds it throughout the pulse, in agreement with recent ultrafast Raman thermometry of underdoped cuprates~\cite{Gatuingt2026Thesis}.
In the weakly correlated metal ($U=0.85$), this mode heats rapidly, exceeding the electronic temperature and reaching a maximum near the center of the pulse. 
Subsequently, its temperature decreases as energy is transferred back to the electronic subsystem. 
Similar behavior is observed in the correlated metallic regime ($U=1.15$), although the transient overheating of the AFM mode becomes considerably more pronounced. 
Near the Mott crossover ($U=1.5$), the initial heating remains comparable to that in the metallic regimes, but the subsequent evolution changes qualitatively. 
As the system approaches the Mott crossover, the AFM temperature first converges toward the electronic temperature, after which the AFM mode undergoes a second pronounced heating stage once the system enters the Mott insulating regime, reflecting the enhanced efficiency of energy transfer into magnetic excitations when itinerant electronic degrees of freedom become suppressed.

The long-wavelength spin mode at $\Gamma$ behaves markedly differently. 
Its effective temperature never exceeds that of the electronic subsystem, indicating substantially weaker energy absorption. 
In the weakly correlated metal [Fig.~\ref{fig:pulse-spectral}\,(g)], it initially follows the rapid heating of the N electrons before gradually approaching the AN electronic temperature. 
This behavior indicates that the spin excitations at $\Gamma$ are primarily coupled to the itinerant electrons and is consistent with the progressive closure of the pseudogap as the electronic system thermalizes. 
The same qualitative behavior persists at ${U=1.15}$ [Fig.~\ref{fig:pulse-spectral}\,(h)], although the effective temperature of the long-wavelength spin excitations remains lower than the electronic temperature during the second half of the pulse. 
Close to the Mott crossover, the heating becomes even less efficient [Fig.~\ref{fig:pulse-spectral}\,(i)], and the spin excitations at the $\Gamma$ point remain colder than the electronic subsystem throughout the accessible time window.

To assess whether the transient overheating of the AFM mode corresponds to a thermal redistribution of magnetic excitations, we compare the nonequilibrium spectral function of magnetic excitations at the M point with equilibrium reference spectra evaluated at the instantaneous electronic temperature and at the effective temperature extracted from the AFM mode itself. 
We analyze two key features of the magnetic spectrum: the peak position, which is controlled by the AFM correlation length, and the peak width, which reflects the damping of the magnetic excitation.

\begin{figure}[t!]
    \centering
    \includegraphics[width=0.98\columnwidth]{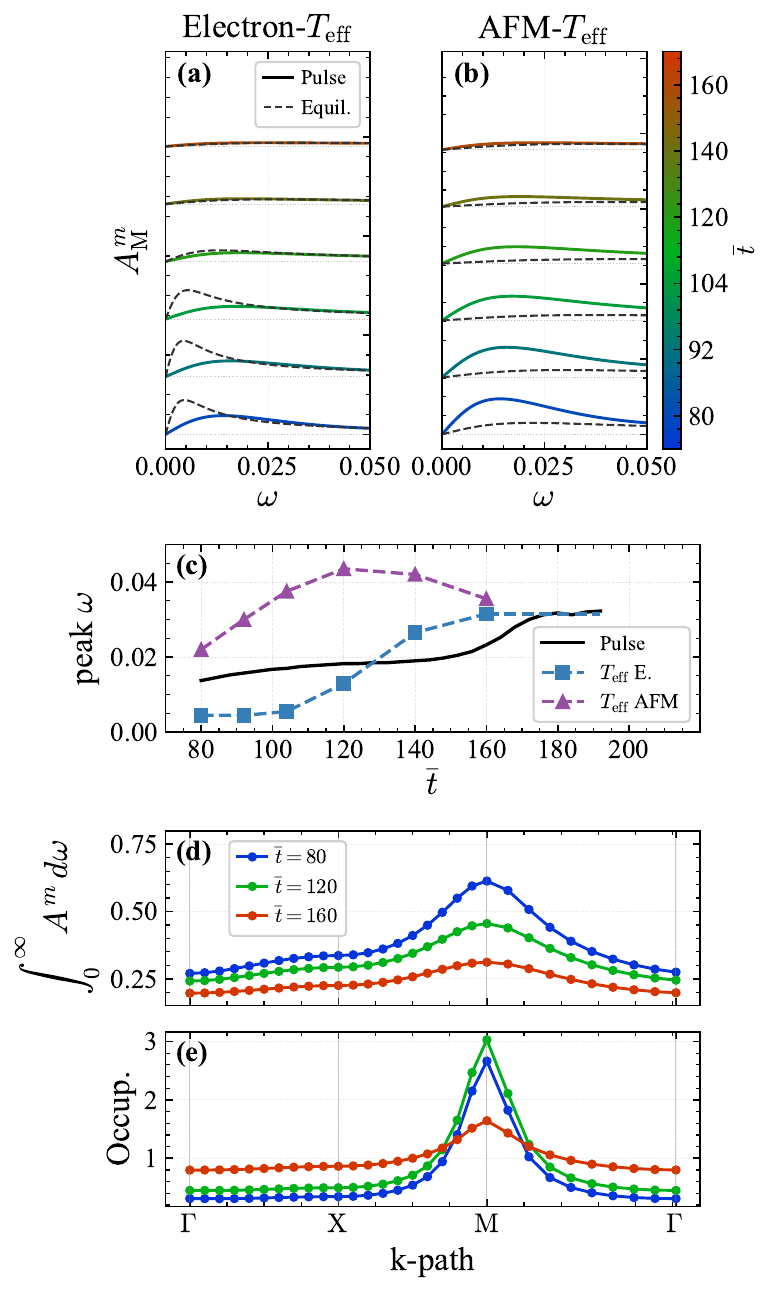}
    \caption{Nonthermal heating and redistribution of the AFM magnon spectral weight for ${U=1.15}$. (a,b)~Nonequilibrium spectral function of magnetic excitations $A^{m}_{\mathrm{M}}(\omega)$ at $\mathrm{M}=(\pi,\pi)$ during the pulse (solid, colored by time $\bar t$) compared with equilibrium references (dashed) evaluated at the instantaneous (a)~average electronic and (b)~AFM mode effective temperatures. Successive times are offset vertically for clarity. (c)~Evolution of the magnon peak position. The nonequilibrium result (\textit{Pulse}, black) is compared with equilibrium peak positions evaluated at the average electronic ($T_\mathrm{eff}$~E.) and AFM magnonic ($T_\mathrm{eff}$~AFM) temperatures. (d)~Momentum-resolved available magnetic spectral weight, $\int_0^\infty A^{m}_{\mathbf q}(\omega)\,d\omega$, along the high-symmetry path $\Gamma$--\,X\,--\,M\,--$\Gamma$ for selected average times $\bar t$. (e)~Corresponding mode occupancy, $\int_0^\infty A^{m,<}_{\mathbf q}(\omega)\,d\omega \big/ \int_0^\infty A^{m}_{\mathbf q}(\omega)\,d\omega$, along the same path. All quantities are in units of the half-bandwidth.}
    \label{fig:magnon_noneq}
\end{figure}

Throughout the pulse, the width of the nonequilibrium spectrum at the M point interpolates between the two equilibrium references: it initially lies between the spectra evaluated at the electronic temperature [Fig.\ref{fig:magnon_noneq}\,(a)] and at the AFM-mode temperature [Fig.\ref{fig:magnon_noneq}\,(b)], while from the middle of the pulse onward (${\bar t=120}$) it follows more closely the spectrum corresponding to the electronic temperature. 
This behavior can be understood from the microscopic damping mechanisms of spin excitations, which are governed by scattering into lower-momentum magnetic fluctuations and by energy transfer to single-particle excitations. 
Since both the electronic subsystem and the long-wavelength spin excitations exhibit similar heating dynamics, as discussed above, these channels provide the dominant contribution to the magnetic excitation lifetime. 
In contrast, the equilibrium spectrum evaluated at the AFM-mode temperature is substantially broader, indicating that a thermal description based solely on the effective AFM temperature overestimates the degree of magnetic incoherence. 
Thus, although the AFM mode reaches a much higher effective temperature than the electrons, the magnetic subsystem cannot be considered internally thermalized during the early stages of the excitation.

The position of the AFM peak in frequency $\omega$, shown by the black curve in Fig.~\ref{fig:magnon_noneq}\,(c), reveals an even stronger deviation from thermal behavior.
While the peak positions of both equilibrium reference spectra, evaluated at the effective electronic (blue curve) and AFM-mode (magenta curve) temperatures, shift to higher energies with increasing temperature, consistent with the equilibrium temperature dependence of magnetic excitations (see, e.g., Ref.~\cite{stepanov_quantum_2018}), the nonequilibrium peak position remains nearly unchanged until ${\bar{t}\approx140}$, approaching the equilibrium value only at later times. 
Thus, although AFM excitations absorb energy almost immediately, the characteristic magnetic correlation length encoded in the AFM peak position remains essentially frozen throughout the pump. The transient dynamics therefore resemble a quench, in which short-wavelength AFM fluctuations initially become weakly coupled to the electronic subsystem and preserve their correlation length before gradually relaxing toward thermal equilibrium.
This mechanism also provides a natural explanation for the second heating stage observed near the Mott crossover [Fig.~\ref{fig:pulse-spectral}\,(i)]. 
As the pump drives the system deeper into the Mott regime, quasiparticle states near the Fermi level become strongly suppressed, reducing the available decay channels for high-energy magnetic fluctuations. 
Consequently, the coupling between spin excitations and electronic degrees of freedom is weakened again, delaying magnetic thermalization despite the continued increase in electronic temperature.

This interpretation is supported by the recent time-resolved RIXS measurements on doped cuprates~\cite{jost2026}, which reveal similarly highly nonthermal dynamics of paramagnons with frozen peak position.
These experiments additionally observe a transient suppression of electron–magnon coupling, inferred from the momentum-dependent redistribution of paramagnon spectral weight, followed by gradual equilibration of the electronic and magnetic subsystems at longer timescales.
Our results demonstrate that this behavior extends to the intermediate-coupling regime. 
Similar behavior has recently been observed in time-resolved Raman measurements of underdoped cuprates~\cite{Gatuingt2026Thesis}, where the effective temperature extracted from the magnetic response rises far above that of the photo-doped carriers, while the two-magnon peak broadens strongly and its energy remains essentially constant.

Further insight into the redistribution of magnetic spectral weight is obtained from the momentum-resolved retarded and lesser spin susceptibilities shown in Fig.~\ref{fig:magnon_noneq}\,(d,\,e). 
Panel~(d) displays the integrated magnetic spectral weight, $\int_0^\infty A^m_{\mathbf{q}}(\omega)\,d\omega$, along the high-symmetry path in momentum space. 
As the pump proceeds, this spectral weight decreases at all momenta, with the strongest suppression occurring at the M point, indicating a progressive weakening of the AFM fluctuations that dominate the equilibrium magnetic response. 
Panel~(e) shows the corresponding mode occupancy, $\int_0^\infty A^{m,<}_{\mathbf{q}}(\omega)\,d\omega \big/ \int_0^\infty A^{m}_{\mathbf{q}}(\omega)\,d\omega$, which characterizes the population of magnetic excitations at different momenta. 
Before the pulse (blue), the occupancy is strongly concentrated near the M point, reflecting the proximity to AFM order. 
Near the middle of the pulse (green), the occupancy increases almost uniformly throughout the BZ. 
The creation of localized doublon–holon pairs by photoexcitation disrupts short-range AFM correlations and generates magnetic excitations with a broad momentum distribution. 
A similarly momentum-independent generation of paramagnons has been proposed in time-resolved RIXS measurements on photoexcited cuprates~\cite{jost2026} and in time-resolved Raman studies of cuprates~\cite{Gatuingt2026Thesis}. 
At later times (red), the occupancy of the AFM mode decreases, while that of lower-momentum modes continues to increase, indicating that AFM excitations redistribute their population toward long-wavelength magnetic fluctuations.

Together, panels~(d) and~(e) reveal a two-stage process.
During the pulse, photoinduced doublon--holon pairs generate magnetic excitations with an almost momentum-independent population: a doublon--holon pair acts as a local defect, and its real-space localization converts into a broad distribution of magnetic excitations across the BZ.
The same mechanism has been proposed to explain the time-resolved Raman measurements on the AFM phase of underdoped cuprates discussed above~\cite{Gatuingt2026Thesis}.
The magnetic subsystem is thus heated directly by the photoexcitation process rather than through the electronic one, placing our results outside the two- and three-temperature description~\cite{Allen1987-three_T, Pankratova2022-three_T}, where the photoexcited carriers relax by emitting magnons and the magnetic temperature is bounded from above by the electronic one at all times.

After the pulse, the high-momentum AFM magnetic excitations undergo a momentum-space cascade, transferring spectral weight toward lower-momentum modes. 
Such a cascade is consistent with the efficient decay of high-energy magnetic excitations inferred for gapless Sr$_2$IrO$_4$, where photoexcited magnons relax through the magnetic dispersion toward low-energy modes~\cite{mazzone2021, dean2016}. 
In contrast, a gap in the magnetic excitation spectrum suppresses this decay channel in gapped antiferromagnets~\cite{mazzone2021}. 
The gapless magnetic spectrum of the present Hubbard model therefore provides favorable conditions for continuous redistribution of magnetic excitation populations toward long wavelengths. 
This mechanism becomes particularly important near the Mott crossover, where the suppression of low-energy quasiparticles weakens the coupling between spin fluctuations and the electronic subsystem, leaving scattering among magnetic excitations as the dominant channel for populating the long-wavelength $\Gamma$ modes.

Despite these distinct transient regimes, the coupled electron–magnetic system exhibits an overall tendency toward thermalization. 
In both metallic regimes, the temperatures of the two magnetic modes gradually converge toward the electronic temperature by the end of the pulse, and the electronic and magnetic subsystems subsequently relax toward a common effective temperature. 
In contrast, near the Mott crossover, relaxation occurs on a much longer timescale and leads to a substantially higher final temperature, consistent with our previous nonequilibrium \textit{D-GW} study~\cite{dasari2025}.

\begin{figure*}[t]
\centering
\includegraphics[width=\textwidth]{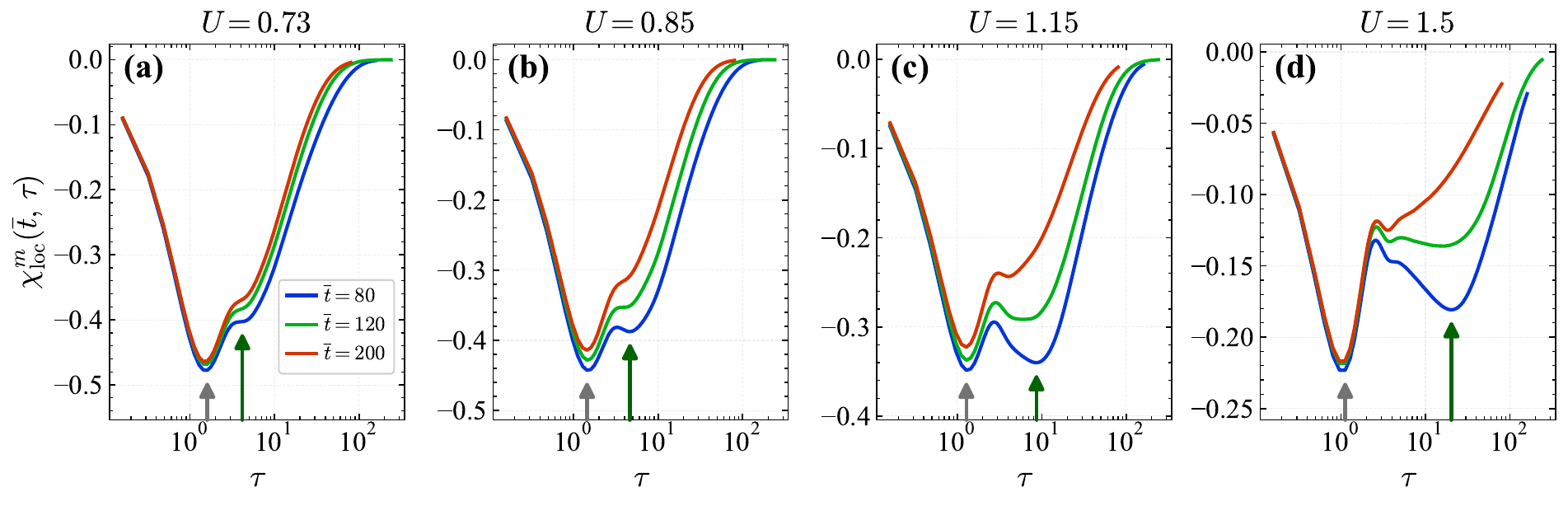}
\caption{Time evolution of the local spin susceptibility $\chi^{m}_{\mathrm{loc}}(\bar t,\tau)$ as a function of the relative time $\tau$ (logarithmic scale) during the pump, for the four interaction strengths $U$ (half-bandwidth units). Curves correspond to increasing average (probe) times $\bar t$. The gray arrow marks the short-time peak ($\tau\sim1/J$), which is independent of $U$, whereas the green arrow marks the slow, interaction-dependent peak that shifts toward larger $\tau$ with increasing $U$ and signals local-moment formation.}
\label{fig:chi-llm}
\end{figure*}

\subsection{Evolution of the local magnetic moment}
\label{sec:llm-criterion}

Having characterized the spin excitations across the most representative interaction strengths, we now turn to the dynamics of the local magnetic moment (LMM). 
To this end, we monitor the local spin susceptibility, $\chi^m_{\rm loc}(\bar{t},\tau)$, resolved as a function of both the average time $\bar{t}$ and the relative time $\tau$, which provides direct access to the intrinsic dynamics of on-site spin fluctuations. Figure~\ref{fig:chi-llm} shows the evolution of $\chi^m_{\rm loc}(\bar{t},\tau)$ as a function of the relative time $\tau$ (shown on a logarithmic scale) for the three interaction strengths considered above, together with an additional calculation at ${U=0.73}$ in the weakly correlated regime. 
The corresponding photoinduced trajectory is indicated by the orange dashed arrow in Fig.~\ref{fig:phase-diagram}\,(a).

For sufficiently strong interactions, the susceptibility exhibits a characteristic two-peak structure. 
The first peak is narrow and located at short times, ${\tau\sim 1/J}$, with a position that is essentially independent of $U$. 
A second, broader peak emerges at longer times and shifts progressively toward larger $\tau$ as $U$ increases, while its spectral weight simultaneously grows. 
This slow component is absent in the weakly correlated regime ($U=0.73$) over the temperature range considered, indicating that it is associated with the formation of long-lived LMMs. 
During the pump excitation, the short-time peak remains nearly unchanged in both position and width. 
In contrast, the long-time peak gradually loses spectral weight and shifts toward shorter times, eventually merging with the short-time contribution, signaling the dynamical melting of the LMM.

To identify the microscopic origin of these peaks, we analyze the half-filled isolated Hubbard dimer in Appendix~\ref{app:dimer}. 
There, we show that the short-time peak originates from bonding–antibonding single-particle excitations, with a characteristic timescale set by the hopping-induced energy splitting, $\sim{1/J}$. 
These processes describe the dynamics of itinerant electrons and contribute to both the local spin and charge susceptibilities. 
In contrast, the slow interaction-dependent peak corresponds to the lattice counterpart of the singlet-to-triplet excitation of the dimer. 
In the strong-coupling limit, its characteristic timescale is governed by the inverse superexchange energy, ${J_{\rm ex}^{-1}\sim U/(4J^2)}$, reflecting the emergence of long-lived LMM dynamics.

Our identification of the slow peak with local-moment dynamics is consistent with Ref.~\cite{sayad_relaxation_2016}, which studied the relaxation of a single classical spin coupled to a one-dimensional Hubbard chain in an external magnetic field. 
By analyzing the local spin susceptibility, the authors identified a two-component structure consisting of a short-time peak, largely independent of $U$ and governed by fast electronic hopping processes, and a slower interaction-dependent contribution associated with the coupling to the LMM. 
Our framework extends this equilibrium analysis of the relaxation of a classical spin to a nonequilibrium framework of correlated electrons, allowing us to resolve the full time evolution of the local spin susceptibility during photoexcitation and to directly monitor the suppression of the LMM contribution in real time. 
We therefore identify the emergence of a well-separated slow peak in $\chi^m_{\rm loc}(\bar{t},\tau)$ as a dynamical signature of the LMM crossover.

The finite width of the slow peak is itself physically significant, as it reflects the finite lifetime of LMM fluctuations. 
These fluctuations are eventually screened at long times through interactions with thermal charge and magnetic fluctuations, Kondo-like screening processes, or spin-flip processes induced by intersite exchange (see, e.g., Refs.~\cite{chalupa_fingerprints_2021, stepanov_spin_2022}). 
The width of the slow peak therefore provides an estimate of the LMM lifetime. 
Characteristic LMM lifetimes in several Hund's metals were quantified in Ref.~\cite{watzenbock_characteristic_2020}, yielding values ranging from a few to several tens of femtoseconds across different material families. 
Using realistic hopping amplitudes for these compounds, ${J\simeq 0.1\text{--}0.5}$\,eV~\cite{Miyake_2010}, we find that the width of our slow peak corresponds to a lifetime of the same order, namely tens of femtoseconds, consistent with these previous estimates.

Since LMM formation is a gradual crossover rather than a sharply defined phase transition, identifying its onset is inherently nontrivial. 
Several complementary diagnostics have therefore been proposed. 
Some rely on fingerprints in the generalized charge susceptibility~\cite{gunnarsson_breakdown_2017, chalupa_fingerprints_2021, mazitov_local_2021}. 
While these approaches successfully identify LMM formation through its impact on charge fluctuations, they are formulated in the charge channel rather than directly in terms of the spin response, and rely on Matsubara-frequency quantities that are not directly accessible experimentally. 
The criterion proposed in Ref.~\cite{stepanov_spin_2022}, which extends the Landau phenomenology of spontaneous symmetry breaking to the LMM crossover, is theoretically more transparent but likewise not directly observable. 
Closer to the present work, Ref.~\cite{gaspard_timescale_2022} extracted a screening timescale from the imaginary-time decay of $\chi^m(\tau)$ within equilibrium DMFT, although this approach remains restricted to imaginary time and captures only local screening processes.

Our criterion combines the advantages of these approaches while extending them to nonequilibrium. 
Formulated directly in the spin channel through the local spin susceptibility $\chi^m_{\rm loc}(\bar t,\tau)$ on the Kadanoff--Baym contour, it provides a real-time measure of LMM dynamics, incorporates the effects of nonlocal magnetic fluctuations through the underlying many-body framework, and resolves the transient melting and recovery of LMM as a function of the average time $\bar{t}$. 
Moreover, because $\chi^m_{\rm loc}(\bar t,\tau)$ represents the local real-time counterpart of the transient dynamical spin response probed by pump–probe RIXS~\cite{jost2026,dean2016,cao2019}, our approach establishes a connection between the microscopic dynamics of LMM formation and experimentally accessible ultrafast spin spectroscopy.

Having established the criterion for LMM formation, we next determine the corresponding boundary throughout the phase diagram. To this end, we perform equilibrium calculations and extract the LMM crossover line as a function of interaction strength and temperature, which is shown in green in Fig.~\ref{fig:phase-diagram},(a).
For the nonequilibrium simulation at $U=0.85$, the melting of the LMM is indicated by the green square along the corresponding heating trajectory. 
As for the other transient crossover points, the melting of the LMM occurs at a higher effective temperature than the equilibrium boundary, further supporting the nonthermal character of the photoinduced ultrafast electron dynamics.

The resulting LMM crossover line is in good agreement with the equilibrium \mbox{D-TRILEX} study~\cite{chatzieleftheriou_local_2024}, where the same crossover was determined using the theoretical LMM formation criterion introduced in Ref.~\cite{stepanov_spin_2022}. 
In particular, we reproduce the characteristic suppression of LMM formation by nonlocal magnetic fluctuations, as well as the convergence of the LMM crossover line toward the AFM transition upon decreasing $U$. 
This behavior is consistent with the Slater regime of magnetism, where LMMs develop as a consequence of the magnetic instability rather than being preformed in the paramagnetic state.
We note, that the quantitative agreement with the equilibrium \mbox{D-TRILEX} results would require the inclusion of three-point vertex corrections and a more accurate solution of the reference DMFT impurity problem.
\section{\label{sec:conclusion}Conclusion}

We have investigated the ultrafast dynamics of the photoexcited half-filled Hubbard model on the square lattice using the real-time \textit{D-GW} method, focusing on the interplay between momentum-selective electronic excitations, collective spin dynamics, and the evolution of local magnetic moments.

We first demonstrated that the equilibrium nodal--antinodal dichotomy leaves a distinct fingerprint in the nonequilibrium electronic response. 
In the metallic and pseudogap regimes, the momentum differentiation of electronic correlations leads to a transient temperature anisotropy across the FS, with N quasiparticles heating more efficiently than AN excitations. 
By comparing the transient spectra with equilibrium spectra evaluated at the same effective temperature, we further revealed that the photoinduced response cannot be described by simple thermal heating. 
Instead, it contains two distinct nonthermal contributions: a universal redistribution of spectral weight around the Fermi level and a correlation-driven transfer of spectral weight from the Hubbard bands into the pseudogap region, producing transient in-gap quasiparticle states characteristic of strong electronic correlations.

The analysis of nonequilibrium spin dynamics revealed a clear separation between high- and low-momentum spin excitations. 
The high-momentum AFM mode absorbs energy rapidly but exhibits a strongly nonthermal response: while its spectral width follows the electronic subsystem from the middle of the pulse onward, its characteristic energy scale and associated correlation length remain nearly frozen during the pump. 
The transient dynamics of AFM fluctuations are therefore reminiscent of a quench, where energy is rapidly injected into the magnetic subsystem while the underlying spatial correlations remain frozen, followed by a slower relaxation toward thermal equilibrium.
In contrast, long-wavelength spin excitations remain more strongly coupled to itinerant quasiparticles and follow the evolution of the electronic subsystem. 
The momentum-resolved spin response further reveals a two-stage relaxation mechanism: photoinduced doublon–holon fluctuations initially generate magnetic excitations throughout the BZ with an almost momentum-independent population, followed by a momentum-space cascade transferring spectral weight from high-momentum AFM fluctuations toward long-wavelength modes. 
Near the Mott crossover, this cascade becomes the dominant pathway for populating low-momentum spin excitations due to the suppression of electron–magnon coupling.

Finally, by analyzing the real-time local spin susceptibility, we identified the slow component of the local spin dynamics as a direct signature of LMM formation. 
This provides a dynamical criterion for tracking the emergence and melting of LMMs during photoexcitation, extending previous equilibrium concepts to nonequilibrium conditions. 
The resulting criterion remains well defined even when conventional spectral signatures of localization become ambiguous at elevated temperatures, highlighting LMM dynamics as a fundamental indicator of correlation-driven localization.

Our results establish real-time \textit{D-GW} as a unified framework capable of simultaneously describing momentum-resolved single-particle excitations, collective spin fluctuations, and LMM dynamics in strongly correlated systems. 
Extending this approach to multi-orbital models and realistic materials represents a natural step toward quantitative simulations of ultrafast correlated-electron phenomena and direct comparison with time-resolved spectroscopic experiments.

\begin{acknowledgments}
The authors thank Martin Eckstein, Michael Potthoff, Igor Krivenko, Yann Gallais, and Laurène Gatuingt for fruitful discussions.
G.B. and E.A.S. acknowledge support from ANR JCJC grant ``ELECTRO'' (ANR-25-CE30-7064).
G.B. also acknowledges support from the Erasmus+ Doctoral Mobility programme of Institut Polytechnique de Paris.
N.D. and A.I.L. acknowledge support from the European Research Council via Synergy Grant No.\ 854843 (the FASTCORR project). 
A.I.L. also acknowledges support from the Deutsche Forschungsgemeinschaft through the research unit QUAST, FOR 5249, Project ID No. 449872909. 
The authors acknowledge the computational resources provided by TGCC-GENCI under project A0180901393 and by the PHYSnet-Rechenzentrum of Universit\"at Hamburg, and we thank Martin Stieben for technical support.
\end{acknowledgments}

\bibliography{biblio}

@article{watzenbock_characteristic_2020,
  title = "{Characteristic Timescales of the Local Moment Dynamics in Hund's Metals}",
  author = {Watzenb\"ock, C. and Edelmann, M. and Springer, D. and Sangiovanni, G. and Toschi, A.},
  journal = {Phys. Rev. Lett.},
  volume = {125},
  issue = {8},
  pages = {086402},
  numpages = {6},
  year = {2020},
  month = {Aug},
  publisher = {American Physical Society},
  doi = {10.1103/PhysRevLett.125.086402},
  url = {https://link.aps.org/doi/10.1103/PhysRevLett.125.086402}
}

@article{sayad_relaxation_2016,
  title = "{Relaxation of a Classical Spin Coupled to a Strongly Correlated Electron System}",
  author = {Sayad, Mohammad and Rausch, Roman and Potthoff, Michael},
  journal = {Phys. Rev. Lett.},
  volume = {117},
  issue = {12},
  pages = {127201},
  numpages = {5},
  year = {2016},
  month = {Sep},
  publisher = {American Physical Society},
  doi = {10.1103/PhysRevLett.117.127201},
  url = {https://link.aps.org/doi/10.1103/PhysRevLett.117.127201}
}

@article{stepanov_spin_2022,
  title = "{Spin dynamics of itinerant electrons: Local magnetic moment formation and Berry phase}",
  author = {Stepanov, E. A. and Brener, S. and Harkov, V. and Katsnelson, M. I. and Lichtenstein, A. I.},
  journal = {Phys. Rev. B},
  volume = {105},
  issue = {15},
  pages = {155151},
  numpages = {21},
  year = {2022},
  month = {Apr},
  publisher = {American Physical Society},
  doi = {10.1103/PhysRevB.105.155151},
  url = {https://link.aps.org/doi/10.1103/PhysRevB.105.155151}
}

@article{chatzieleftheriou_local_2024,
  title = "{Local and Nonlocal Electronic Correlations at the Metal-Insulator Transition in the Two-Dimensional Hubbard Model}",
  author = {Chatzieleftheriou, Maria and Biermann, Silke and Stepanov, Evgeny A.},
  journal = {Phys. Rev. Lett.},
  volume = {132},
  issue = {23},
  pages = {236504},
  numpages = {7},
  year = {2024},
  month = {Jun},
  publisher = {American Physical Society},
  doi = {10.1103/PhysRevLett.132.236504},
  url = {https://link.aps.org/doi/10.1103/PhysRevLett.132.236504}
}

@article{stepanov_quantum_2018,
  title = "{Quantum spin fluctuations and evolution of electronic structure in cuprates}",
  author = {Stepanov, Evgeny A. and Peters, Lars and Krivenko, Igor S. and Lichtenstein, Alexander I. and Katsnelson, Mikhail I. and Rubtsov, Alexey N.},
  journal = {npj Quantum Mater.},
  volume = {3},
  pages = {54},
  numpages = {8},
  year = {2018},
  month = {Oct},
  publisher = {Nature Publishing Group},
  doi = {10.1038/s41535-018-0128-x},
  url = {https://www.nature.com/articles/s41535-018-0128-x}
}

@article{peierls1933,
  title = "{On the Theory of the Diamagnetism of Conduction Electrons}",
  author = {Peierls, R.},
  journal = {Z. Phys.},
  volume = {80},
  number = {11--12},
  pages = {763--791},
  numpages = {29},
  year = {1933},
  doi = {10.1007/BF01342591}
}

@article{dasari2025,
  title = "{Electron-magnon dynamics triggered by an ultrashort laser pulse: A real-time dual $GW$ study}",
  author = {Dasari, Nagamalleswararao and Strand, Hugo U. R. and Eckstein, Martin and Lichtenstein, Alexander I. and Stepanov, Evgeny A.},
  journal = {Phys. Rev. B},
  volume = {111},
  issue = {23},
  pages = {235129},
  numpages = {20},
  year = {2025},
  month = {Jun},
  publisher = {American Physical Society},
  doi = {10.1103/vglv-2rmv},
  url = {https://link.aps.org/doi/10.1103/vglv-2rmv}
}

@article{Stepanov2019,
  title = "{Consistent partial bosonization of the extended Hubbard model}",
  author = {Stepanov, E. A. and Harkov, V. and Lichtenstein, A. I.},
  journal = {Phys. Rev. B},
  volume = {100},
  issue = {20},
  pages = {205115},
  numpages = {13},
  year = {2019},
  month = {Nov},
  publisher = {American Physical Society},
  doi = {10.1103/PhysRevB.100.205115},
  url = {https://link.aps.org/doi/10.1103/PhysRevB.100.205115}
}

@article{stepanov2021_1,
  title = {Impact of partially bosonized collective fluctuations on electronic degrees of freedom},
  author = {Harkov, V. and Vandelli, M. and Brener, S. and Lichtenstein, A. I. and Stepanov, E. A.},
  journal = {Phys. Rev. B},
  volume = {103},
  issue = {24},
  pages = {245123},
  numpages = {18},
  year = {2021},
  month = {Jun},
  publisher = {American Physical Society},
  doi = {10.1103/PhysRevB.103.245123},
  url = {https://link.aps.org/doi/10.1103/PhysRevB.103.245123}
}

@article{stepanov2021_2,
	title = "{Multi-band D-TRILEX approach to materials with strong electronic correlations}",
	pages = {036},
	author = {Vandelli, Matteo and Kaufmann, Josef and El-Nabulsi, Mohammed and Harkov, Viktor and Lichtenstein, Alexander and Stepanov, Evgeny},
	journal = {SciPost Phys.},
	volume = {13},
	year = {2022},
	publisher = {SciPost},
	doi = {10.21468/SciPostPhys.13.2.036},
	url = {https://scipost.org/10.21468/SciPostPhys.13.2.036}
}

@phdthesis{StepanovHDR,
      title="{Diagrammatics in the Dual Space, or There and Back Again}",
      author={Evgeny A. Stepanov},
      school = {{\'E}cole Polytechnique},
      year={2025},
      type = {Habilitation thesis},
      doi = {10.48550/arXiv.2512.12389},
      url = {https://arxiv.org/abs/2512.12389}
}

@article{eckstein2010,
  title = "{Nonequilibrium dynamical mean-field calculations based on the noncrossing approximation and its generalizations}",
  author = {Eckstein, Martin and Werner, Philipp},
  journal = {Phys. Rev. B},
  volume = {82},
  issue = {11},
  pages = {115115},
  numpages = {13},
  year = {2010},
  month = {Sep},
  publisher = {American Physical Society},
  doi = {10.1103/PhysRevB.82.115115},
  url = {https://link.aps.org/doi/10.1103/PhysRevB.82.115115}
}

@article{schuler2020,
  author  = {Sch\"uler, Michael and Gole\v{z}, Denis and Murakami, Yuta and Bittner, Nikolaj and Herrmann, Andreas and Strand, Hugo U. R. and Werner, Philipp and Eckstein, Martin},
  title   = "{NESSi: The Non-Equilibrium Systems Simulation package}",
  journal = {Comput. Phys. Commun.},
  volume  = {257},
  pages   = {107484},
  year    = {2020},
  doi     = {10.1016/j.cpc.2020.107484},
}

@article{dasari2018,
  title = "{Photoexcited states in correlated band insulators}",
  author = {Dasari, Nagamalleswararao and Eckstein, Martin},
  journal = {Phys. Rev. B},
  volume = {98},
  issue = {3},
  pages = {035113},
  numpages = {9},
  year = {2018},
  month = {Jul},
  publisher = {American Physical Society},
  doi = {10.1103/PhysRevB.98.035113},
  url = {https://link.aps.org/doi/10.1103/PhysRevB.98.035113}
}

@article{murakami2023,
  title = "{Photoinduced nonequilibrium states in Mott insulators}",
  author = {Murakami, Yuta and Gole\ifmmode \check{z}\else \v{z}\fi{}, Denis and Eckstein, Martin and Werner, Philipp},
  journal = {Rev. Mod. Phys.},
  volume = {97},
  issue = {3},
  pages = {035001},
  numpages = {63},
  year = {2025},
  month = {Jul},
  publisher = {American Physical Society},
  doi = {10.1103/tkjh-lr83},
  url = {https://link.aps.org/doi/10.1103/tkjh-lr83}
}

@article{aoki2014,
  author  = {Aoki, Hideo and Tsuji, Naoto and Eckstein, Martin and Kollar, Marcus and Oka, Takashi and Werner, Philipp},
  title   = "{Nonequilibrium dynamical mean-field theory and its applications}",
  journal = {Rev. Mod. Phys.},
  volume  = {86},
  pages   = {779--837},
  year    = {2014},
  doi     = {10.1103/RevModPhys.86.779},
}

@article{fausti2011,
  title = "{Light-Induced Superconductivity in a Stripe-Ordered Cuprate}",
  author = {Fausti, D. and Tobey, R. I. and Dean, N. and Kaiser, S. and Dienst, A. and Hoffmann, M. C. and Pyon, S. and Takayama, T. and Takagi, H. and Cavalleri, A.},
  journal = {Science},
  volume = {331},
  number = {6014},
  pages = {189--191},
  numpages = {3},
  year = {2011},
  month = {Jan},
  publisher = {American Association for the Advancement of Science},
  doi = {10.1126/science.1197294}
}

@article{delatorre2021,
  title = "{Colloquium: Nonthermal pathways to ultrafast control in quantum materials}",
  author = {de la Torre, Alberto and Kennes, Dante M. and Claassen, Martin and Gerber, Simon and McIver, James W. and Sentef, Michael A.},
  journal = {Rev. Mod. Phys.},
  volume = {93},
  number = {4},
  pages = {041002},
  numpages = {31},
  year = {2021},
  month = {Oct},
  doi = {10.1103/RevModPhys.93.041002}
}

@article{stojchevska2014,
  title = "{Ultrafast Switching to a Stable Hidden Quantum State in an Electronic Crystal}",
  author = {Stojchevska, L. and Vaskivskyi, I. and Mertelj, T. and Kusar, P. and Svetin, D. and Brazovskii, S. and Mihailovic, D.},
  journal = {Science},
  volume = {344},
  number = {6180},
  pages = {177--180},
  numpages = {4},
  year = {2014},
  month = {Apr},
  publisher = {American Association for the Advancement of Science},
  doi = {10.1126/science.1241591}
}

@article{basov2017,
  title = "{Towards properties on demand in quantum materials}",
  author = {Basov, D. N. and Averitt, R. D. and Hsieh, D.},
  journal = {Nat. Mater.},
  volume = {16},
  number = {11},
  pages = {1077--1088},
  numpages = {12},
  year = {2017},
  month = {Oct},
  publisher = {Nature Publishing Group},
  doi = {10.1038/nmat5017}
}

@article{oka2019,
  title = "{Floquet Engineering of Quantum Materials}",
  author = {Oka, Takashi and Kitamura, Sota},
  journal = {Annu. Rev. Condens. Matter Phys.},
  volume = {10},
  pages = {387--408},
  numpages = {22},
  year = {2019},
  month = {Mar},
  doi = {10.1146/annurev-conmatphys-031218-013423}
}

@article{damascelli2003,
  title = "{Angle-resolved photoemission studies of the cuprate superconductors}",
  author = {Damascelli, Andrea and Hussain, Zahid and Shen, Zhi-Xun},
  journal = {Rev. Mod. Phys.},
  volume = {75},
  number = {2},
  pages = {473--541},
  numpages = {69},
  year = {2003},
  month = {Apr},
  doi = {10.1103/RevModPhys.75.473}
}

@article{sobota2021,
  title = "{Angle-resolved photoemission studies of quantum materials}",
  author = {Sobota, Jonathan A. and He, Yu and Shen, Zhi-Xun},
  journal = {Rev. Mod. Phys.},
  volume = {93},
  number = {2},
  pages = {025006},
  year = {2021},
  month = {May},
  doi = {10.1103/RevModPhys.93.025006}
}

@article{dasari_nonlocal_2025,
  title = "{Nonlocal Correlation Effects in dc and Optical Conductivity of the Hubbard Model}",
  author = {Dasari, Nagamalleswararao and Strand, Hugo U. R. and Eckstein, Martin and Lichtenstein, Alexander I. and Stepanov, Evgeny A.},
  journal = {Phys. Rev. Lett.},
  volume = {136},
  issue = {10},
  pages = {106905},
  numpages = {7},
  year = {2026},
  month = {Mar},
  publisher = {American Physical Society},
  doi = {10.1103/gzwp-zd6t},
  url = {https://link.aps.org/doi/10.1103/gzwp-zd6t}
}

@article{cilento2018photoenhanced,
  title   = "{Dynamics of correlation-frozen antinodal quasiparticles in superconducting cuprates}",
  author  = {Cilento, F. and Manzoni, G. and Sterzi, A. and Peli, S. and Ronchi, A. and Crepaldi, A. and Boschini, F. and Cacho, C. and Chapman, R. and Springate, E. and Eisaki, H. and Greven, M. and Berciu, M. and Kemper, A. F. and Damascelli, A. and Capone, M. and Giannetti, C. and Parmigiani, F.},
  journal = {Sci. Adv.},
  volume  = {4},
  number  = {2},
  pages   = {eaar1998},
  year    = {2018},
  month   = {Feb},
  doi     = {10.1126/sciadv.aar1998}
}

@article{cilento2014,
  title   = "{Photo-enhanced antinodal conductivity in the pseudogap state of high-$T_c$ cuprates}",
  author  = {Cilento, F. and Dal Conte, S. and Coslovich, G. and Peli, S. and Nembrini, N. and Mor, S. and Banfi, F. and Ferrini, G. and Eisaki, H. and Chan, M. K. and Dorow, C. J. and Veit, M. J. and Greven, M. and van der Marel, D. and Comin, R. and Damascelli, A. and Rettig, L. and Bovensiepen, U. and Capone, M. and Giannetti, C. and Parmigiani, F.},
  journal = {Nat. Commun.},
  volume  = {5},
  pages   = {4353},
  year    = {2014},
  month   = {Jul},
  publisher = {Nature Publishing Group},
  doi     = {10.1038/ncomms5353}
}

@article{zonno2021,
  title = "{Time-resolved ARPES on cuprates: Tracking the low-energy electrodynamics in the time domain}",
  author = {Zonno, M. and Boschini, F. and Damascelli, A.},
  journal = {J. Electron Spectrosc. Relat. Phenom.},
  volume = {251},
  pages = {147091},
  year = {2021},
  issn = {0368-2048},
  doi = {10.1016/j.elspec.2021.147091},
  url = {https://www.sciencedirect.com/science/article/pii/S0368204821000463}
}

@article{mazzone2021,
  title   = "{Laser-induced transient magnons in Sr$_3$Ir$_2$O$_7$ throughout the Brillouin zone}",
  author  = {Mazzone, D. G. and Meyers, D. and Cao, Y. and Vale, J. G. and Dashwood, C. D. and Shi, Y. and James, A. J. A. and Robinson, N. J. and Lin, J. and Thampy, V. and Tanaka, Y. and Johnson, A. S. and Miao, H. and Wang, R. and Assefa, T. A. and Kim, J. and Casa, D. and Mankowsky, R. and Zhu, D. and Alonso-Mori, R. and Song, S. and Yavas, H. and Katayama, T. and Yabashi, M. and Kubota, Y. and Owada, S. and Liu, J. and Yang, J. and Konik, R. M. and Robinson, I. K. and Hill, J. P. and McMorrow, D. F. and F{\"o}rst, M. and Wall, S. and Liu, X. and Dean, M. P. M.},
  journal = {Proc. Natl. Acad. Sci. U.S.A.},
  volume  = {118},
  number  = {22},
  pages   = {e2103696118},
  year    = {2021},
  month   = {Jun},
  doi     = {10.1073/pnas.2103696118}
}

@article{dean2016,
  title = "{Ultrafast energy- and momentum-resolved dynamics of magnetic correlations in the photo-doped Mott insulator Sr$_2$IrO$_4$}",
  author = {Dean, M. P. M. and Cao, Y. and Liu, X. and Wall, S. and Zhu, D. and Mankowsky, R. and Thampy, V. and Chen, X. M. and Vale, J. G. and Casa, D. and Kim, Jungho and Said, A. H. and Juhas, P. and Alonso-Mori, R. and Glownia, J. M. and Robert, A. and Robinson, J. and Sikorski, M. and Song, S. and Kozina, M. and Lemke, H. and Patthey, L. and Owada, S. and Katayama, T. and Yabashi, M. and Tanaka, Yoshikazu and Togashi, T. and Liu, J. and Rayan Serrao, C. and Kim, B. J. and Huber, L. and Chang, C.-L. and McMorrow, D. F. and F{\"o}rst, M. and Hill, J. P.},
  journal = {Nat. Mater.},
  volume = {15},
  number = {6},
  pages = {601--605},
  year = {2016},
  month = {Jun},
  doi = {10.1038/nmat4641}
}

@article{cao2019,
  title = "{Ultrafast dynamics of spin and orbital correlations in quantum materials: an energy- and momentum-resolved perspective}",
  author = {Cao, Y. and Mazzone, D. G. and Meyers, D. and Hill, J. P. and Liu, X. and Wall, S. and Dean, M. P. M.},
  journal = {Philos. Trans. R. Soc. A},
  volume = {377},
  number = {2145},
  pages = {20170480},
  year = {2019},
  month = {May},
  doi = {10.1098/rsta.2017.0480}
}

@article{afanasiev2019,
  title = "{Ultrafast Spin Dynamics in Photodoped Spin-Orbit Mott Insulator ${\mathrm{Sr}}_{2}{\mathrm{IrO}}_{4}$}",
  author = {Afanasiev, D. and Gatilova, A. and Groenendijk, D. J. and Ivanov, B. A. and Gibert, M. and Gariglio, S. and Mentink, J. and Li, J. and Dasari, N. and Eckstein, M. and Rasing, Th. and Caviglia, A. D. and Kimel, A. V.},
  journal = {Phys. Rev. X},
  volume = {9},
  issue = {2},
  pages = {021020},
  numpages = {13},
  year = {2019},
  month = {Apr},
  publisher = {American Physical Society},
  doi = {10.1103/PhysRevX.9.021020},
  url = {https://link.aps.org/doi/10.1103/PhysRevX.9.021020}
}

@article{Iwai2003,
  title = "{Ultrafast Optical Switching to a Metallic State by Photoinduced Mott Transition in a Halogen-Bridged Nickel-Chain Compound}",
  author = {Iwai, S. and Ono, M. and Maeda, A. and Matsuzaki, H. and Kishida, H. and Okamoto, H. and Tokura, Y.},
  journal = {Phys. Rev. Lett.},
  volume = {91},
  issue = {5},
  pages = {057401},
  numpages = {4},
  year = {2003},
  month = {Jul},
  publisher = {American Physical Society},
  doi = {10.1103/PhysRevLett.91.057401},
  url = {https://link.aps.org/doi/10.1103/PhysRevLett.91.057401}
}

@misc{schmidt2002NeqDMFT,
      title="{Nonequilibrium dynamical mean-field theory of a strongly correlated system}", 
      author={P. Schmidt and H. Monien},
      year={2002},
      eprint={cond-mat/0202046},
      archivePrefix={arXiv},
      primaryClass={cond-mat.str-el},
      url={https://arxiv.org/abs/cond-mat/0202046}, 
}

@article{slater1951,
  title = "{Magnetic Effects and the Hartree-Fock Equation}",
  author = {Slater, J. C.},
  journal = {Phys. Rev.},
  volume = {82},
  issue = {4},
  pages = {538--541},
  numpages = {0},
  year = {1951},
  month = {May},
  publisher = {American Physical Society},
  doi = {10.1103/PhysRev.82.538},
  url = {https://link.aps.org/doi/10.1103/PhysRev.82.538}
}

@article{rohringer2016,
  title = "{Impact of nonlocal correlations over different energy scales: A dynamical vertex approximation study}",
  author = {Rohringer, G. and Toschi, A.},
  journal = {Phys. Rev. B},
  volume = {94},
  issue = {12},
  pages = {125144},
  numpages = {30},
  year = {2016},
  month = {Sep},
  publisher = {American Physical Society},
  doi = {10.1103/PhysRevB.94.125144},
  url = {https://link.aps.org/doi/10.1103/PhysRevB.94.125144}
}

@article{anderson1959,
  title = "{New Approach to the Theory of Superexchange Interactions}",
  author = {Anderson, P. W.},
  journal = {Phys. Rev.},
  volume = {115},
  issue = {1},
  pages = {2--13},
  numpages = {0},
  year = {1959},
  month = {Jul},
  publisher = {American Physical Society},
  doi = {10.1103/PhysRev.115.2},
  url = {https://link.aps.org/doi/10.1103/PhysRev.115.2}
}

@article{chao1977degenerate,
  author  = {Chao, K. A. and Spa{\l}ek, J. and Ole{\'s}, A. M.},
  title   = "{Degenerate Perturbation Theory and Its Application to the Hubbard Model}",
  journal = {Phys. Lett. A},
  volume  = {64},
  number  = {2},
  pages   = {163--166},
  year    = {1977},
  month   = {Dec},
  doi     = {10.1016/0375-9601(77)90702-2}
}

@article{chao1977kinetic,
  author  = {Chao, K. A. and Spa{\l}ek, J. and Ole{\'s}, A. M.},
  title   = "{Kinetic Exchange Interaction in a Narrow S-Band}",
  journal = {J. Phys. C},
  volume  = {10},
  number  = {10},
  pages   = {L271--L276},
  year    = {1977},
  doi     = {10.1088/0022-3719/10/10/002}
}

@article{macdonald1988,
  title = "{$\frac{t}{U}$ expansion for the Hubbard model}",
  author = {MacDonald, A. H. and Girvin, S. M. and Yoshioka, D.},
  journal = {Phys. Rev. B},
  volume = {37},
  issue = {16},
  pages = {9753--9756},
  numpages = {0},
  year = {1988},
  month = {Jun},
  publisher = {American Physical Society},
  doi = {10.1103/PhysRevB.37.9753},
  url = {https://link.aps.org/doi/10.1103/PhysRevB.37.9753}
}

@article{spalek2007,
  author  = {Spa{\l}ek, J.},
  title   = "{$t$--$J$ Model Then and Now: A Personal Perspective from the Pioneering Times}",
  journal = {Acta Phys. Pol. A},
  volume  = {111},
  pages   = {409--424},
  year    = {2007},
  doi     = {10.12693/APhysPolA.111.409}
}

@article{mazitov_local_2021,
  title = "{Local magnetic moment formation and Kondo screening in the half-filled single-band Hubbard model}",
  author = {Mazitov, T. B. and Katanin, A. A.},
  journal = {Phys. Rev. B},
  volume = {105},
  issue = {8},
  pages = {L081111},
  numpages = {6},
  year = {2022},
  month = {Feb},
  publisher = {American Physical Society},
  doi = {10.1103/PhysRevB.105.L081111},
  url = {https://link.aps.org/doi/10.1103/PhysRevB.105.L081111}
}

@article{mazitov_local_2024,
  title = "{Local magnetic moment formation and Kondo screening in the presence of Hund exchange: Two-band Hubbard model analysis}",
  author = {Mazitov, T. B. and Katanin, A. A.},
  journal = {Phys. Rev. B},
  volume = {110},
  issue = {7},
  pages = {075160},
  numpages = {11},
  year = {2024},
  month = {Aug},
  publisher = {American Physical Society},
  doi = {10.1103/PhysRevB.110.075160},
  url = {https://link.aps.org/doi/10.1103/PhysRevB.110.075160}
}

@misc{realevol,
  author       = {Krivenko, Igor and Danilov, Mikhail and Kubiczek, Piotr},
  title        = "{realevol: Real time evolution solver based on {TRIQS}}",
  year         = {2026},
  note         = {Version 0.11.2},
  howpublished = {\url{https://github.com/krivenko/triqs-realevol}},
  publisher = {GitHub}
}

@article{TRIQS2015,
  author  = {Parcollet, Olivier and Ferrero, Michel and Ayral, Thomas and Hafermann, Hartmut and Krivenko, Igor and Messio, Laura and Seth, Priyanka},
  title   = "{TRIQS: A toolbox for research on interacting quantum systems}",
  journal = {Comput. Phys. Commun.},
  volume  = {196},
  pages   = {398--415},
  year    = {2015},
  month   = {Nov},
  doi     = {10.1016/j.cpc.2015.04.023},
}

@article{Miyake_2010,
  author  = {Miyake, Takashi and Nakamura, Kazuma and Arita, Ryotaro and Imada, Masatoshi},
  title   = "{Comparison of Ab initio Low-Energy Models for LaFePO, LaFeAsO, BaFe$_2$As$_2$, LiFeAs, FeSe, and FeTe: Electron Correlation and Covalency}",
  journal = {J. Phys. Soc. Jpn.},
  volume  = {79},
  number  = {4},
  pages   = {044705},
  year    = {2010},
  month   = {Apr},
  doi     = {10.1143/JPSJ.79.044705}
}

@article{carrascal2015hubbard,
  author  = {Carrascal, D. J. and Ferrer, J. and Smith, J. C. and Burke, K.},
  title   = "{The Hubbard dimer: a density functional case study of a many-body problem}",
  journal = {J. Phys.: Condens. Matter},
  volume  = {27},
  number  = {39},
  pages   = {393001},
  year    = {2015},
  doi     = {10.1088/0953-8984/27/39/393001}
}

@article{Avella2004,
  author  = {Avella, A. and Mancini, F. and Saikawa, T.},
  title   = "{The 2-site Hubbard and $t$-$J$ models}",
  journal = {Eur. Phys. J. B},
  volume  = {36},
  pages   = {445--473},
  year    = {2003},
  doi     = {10.1140/epjb/e2004-00002-8},
}

@article{gunnarsson_breakdown_2017,
  title = {Breakdown of Traditional Many-Body Theories for Correlated Electrons},
  author = {Gunnarsson, O. and Rohringer, G. and Sch\"afer, T. and Sangiovanni, G. and Toschi, A.},
  journal = {Phys. Rev. Lett.},
  volume = {119},
  issue = {5},
  pages = {056402},
  numpages = {5},
  year = {2017},
  month = {Aug},
  publisher = {American Physical Society},
  doi = {10.1103/PhysRevLett.119.056402},
  url = {https://link.aps.org/doi/10.1103/PhysRevLett.119.056402}
}

@article{chalupa_fingerprints_2021,
  title = "{Fingerprints of the Local Moment Formation and its Kondo Screening in the Generalized Susceptibilities of Many-Electron Problems}",
  author = {Chalupa, P. and Sch\"afer, T. and Reitner, M. and Springer, D. and Andergassen, S. and Toschi, A.},
  journal = {Phys. Rev. Lett.},
  volume = {126},
  issue = {5},
  pages = {056403},
  numpages = {9},
  year = {2021},
  month = {Feb},
  publisher = {American Physical Society},
  doi = {10.1103/PhysRevLett.126.056403},
  url = {https://link.aps.org/doi/10.1103/PhysRevLett.126.056403}
}

@article{gaspard_timescale_2022,
  title   = "{Timescale of Local Moment Screening across and above the Mott Transition}",
  author  = {Gaspard, L{\'e}o and Tomczak, Jan M.},
  journal = {SciPost Phys.},
  volume  = {12},
  pages   = {184},
  year    = {2022},
  doi     = {10.21468/SciPostPhys.12.6.184},
  archivePrefix = {arXiv},
  eprint  = {2112.02881}
}

@inproceedings{Gallais2026,
  title = "{Tracking photo-induced superconducting to normal state transition in the cuprate Bi2Sr2CaCu2O8}",
  author = {Gallais, Yann},
  booktitle = {Advances in Ultrafast Condensed Phase Physics V},
  editor = {Haacke, Stefan and Ossiander, Markus},
  volume = {PC14077},
  pages = {PC1407709},
  organization = {International Society for Optics and Photonics},
  publisher = {SPIE},
  year = {2026},
  doi = {10.1117/12.3109486},
  url = {https://doi.org/10.1117/12.3109486}
}

@article{Gatuingt2026,
  title = "{Ultrafast Raman probe of the photoinduced superconducting to normal state transition in the cuprate ${\mathrm{Bi}}_{2}{\mathrm{Sr}}_{2}{\mathrm{CaCu}}_{2}{\mathrm{O}}_{8+\ensuremath{\delta}}$}",
  author = {Gatuingt, Laur\`ene and Alekhin, Alexandr and Nilforoushan, Niloufar and Houver, Sarah and Sacuto, Alain and Gu, Genda and Gallais, Yann},
  journal = {Phys. Rev. B},
  volume = {113},
  issue = {1},
  pages = {014509},
  numpages = {12},
  year = {2026},
  month = {Jan},
  publisher = {American Physical Society},
  doi = {10.1103/wkx7-lcqf},
  url = {https://link.aps.org/doi/10.1103/wkx7-lcqf}
}

@article{Katsumi2025THzRaman,
  title = "{Distinct terahertz nonlinear and Raman responses in cuprate superconductors {Bi$_2$Sr$_2$CaCu$_2$O$_{8+x}$}}",
  author = {Katsumi, Kota and Gallais, Yann and Shimano, Ryo},
  journal = {npj Quantum Mater.},
  volume = {10},
  pages = {91},
  year = {2025},
  publisher = {Springer Nature},
  doi = {10.1038/s41535-025-00807-x},
  url = {https://doi.org/10.1038/s41535-025-00807-x}
}

@article{kanigel2006,
  title = "{Evolution of the pseudogap from Fermi arcs to the nodal liquid}",
  author = {Kanigel, A. and Norman, M. R. and Randeria, M. and Chatterjee, U. and Souma, S. and Kaminski, A. and Fretwell, H. M. and Rosenkranz, S. and Shi, M. and Sato, T. and Takahashi, T. and Li, Z. Z. and Raffy, H. and Kadowaki, K. and Hinks, D. and Ozyuzer, L. and Campuzano, J. C.},
  journal = {Nat. Phys.},
  volume = {2},
  pages = {447--451},
  year = {2006},
  doi = {10.1038/nphys334}
}

@article{norman1998,
  title = "{Destruction of the Fermi surface in underdoped high-$T_c$ superconductors}",
  author = {Norman, M. R. and Ding, H. and Randeria, M. and Campuzano, J. C. and Yokoya, T. and Takeuchi, T. and Takahashi, T. and Mochiku, T. and Kadowaki, K. and Guptasarma, P. and Hinks, D. G.},
  journal = {Nature},
  volume = {392},
  pages = {157--160},
  year = {1998},
  doi = {10.1038/32366}
}

@article{jost2026,
  title   = "{Collective magnetic excitations in a photoexcited electron-doped cuprate superconductor}",
  author  = {Jost, Daniel and Li, Jiarui and Hales, Jordyn and Sobota, Jonathan and Merzoni, Giacomo and Martinelli, Leonardo and Ding, Shuhan and Xu, Ke-Jun and Schlappa, Justine and Scherz, Andreas and Carley, Robert and Van Kuiken, Benjamin E. and Asmara, Teguh C. and Hoang, Le Phuong and Mercadier, Laurent and Parchenko, Sergii and Teichmann, Martin and Kirchmann, Patrick S. and Ghiringhelli, Giacomo and Moritz, Brian and Shen, Zhi-Xun and Devereaux, Thomas P. and Wang, Yao and Lee, Wei-Sheng},
  journal = {Phys. Rev. Lett.},
  year    = {2026},
  doi     = {10.1103/2h9x-8tjk}
}

@misc{Gatuingt2026Thesis,
  author = {Gatuingt, Laur{\`e}ne and Gallais, Yann},
  note   = {private communication}
}

@article{mitrano2020trrixs,
  title = "{Probing light-driven quantum materials with ultrafast resonant inelastic X-ray scattering}",
  author = {Mitrano, Matteo and Wang, Yao},
  journal = {Commun. Phys.},
  volume = {3},
  pages = {184},
  year = {2020},
  month = {Oct},
  publisher = {Nature Publishing Group},
  doi = {10.1038/s42005-020-00447-6}
}

@article{Parham2017,
  title = "{Ultrafast Gap Dynamics and Electronic Interactions in a Photoexcited Cuprate Superconductor}",
  author = {Parham, S. and Li, H. and Nummy, T. J. and Waugh, J. A. and Zhou, X. Q. and Griffith, J. and Schneeloch, J. and Zhong, R. D. and Gu, G. D. and Dessau, D. S.},
  journal = {Phys. Rev. X},
  volume = {7},
  issue = {4},
  pages = {041013},
  numpages = {11},
  year = {2017},
  month = {Oct},
  publisher = {American Physical Society},
  doi = {10.1103/PhysRevX.7.041013},
  url = {https://link.aps.org/doi/10.1103/PhysRevX.7.041013}
}

@article{Hashimoto2014,
  author = {Hashimoto, Makoto and Vishik, Inna M. and He, Rui-Hua and Devereaux, Thomas P. and Shen, Zhi-Xun},
  title = "{Energy gaps in high-transition-temperature cuprate superconductors}",
  journal = {Nat. Phys.},
  volume = {10},
  number = {7},
  pages = {483--495},
  year = {2014},
  month = {Jul},
  publisher = {Nature Publishing Group},
  doi = {10.1038/nphys3009}
}

@article{Pankratova2022-three_T,
  title = "{Heat-conserving three-temperature model for ultrafast demagnetization in nickel}",
  author = {Pankratova, M. and Miranda, I. P. and Thonig, D. and Pereiro, M. and Sj\"oqvist, E. and Delin, A. and Eriksson, O. and Bergman, A.},
  journal = {Phys. Rev. B},
  volume = {106},
  issue = {17},
  pages = {174407},
  numpages = {10},
  year = {2022},
  month = {Nov},
  publisher = {American Physical Society},
  doi = {10.1103/PhysRevB.106.174407},
  url = {https://link.aps.org/doi/10.1103/PhysRevB.106.174407}
}

@article{Allen1987-three_T,
  title = "{Theory of thermal relaxation of electrons in metals}",
  author = {Allen, Philip B.},
  journal = {Phys. Rev. Lett.},
  volume = {59},
  issue = {13},
  pages = {1460--1463},
  numpages = {0},
  year = {1987},
  month = {Sep},
  publisher = {American Physical Society},
  doi = {10.1103/PhysRevLett.59.1460},
  url = {https://link.aps.org/doi/10.1103/PhysRevLett.59.1460}
}

\clearpage

\appendix

\section{Hubbard dimer analysis of the spin susceptibility peaks}
\label{app:dimer}

\begin{figure*}[ht]
\centering
\includegraphics[width=\textwidth]{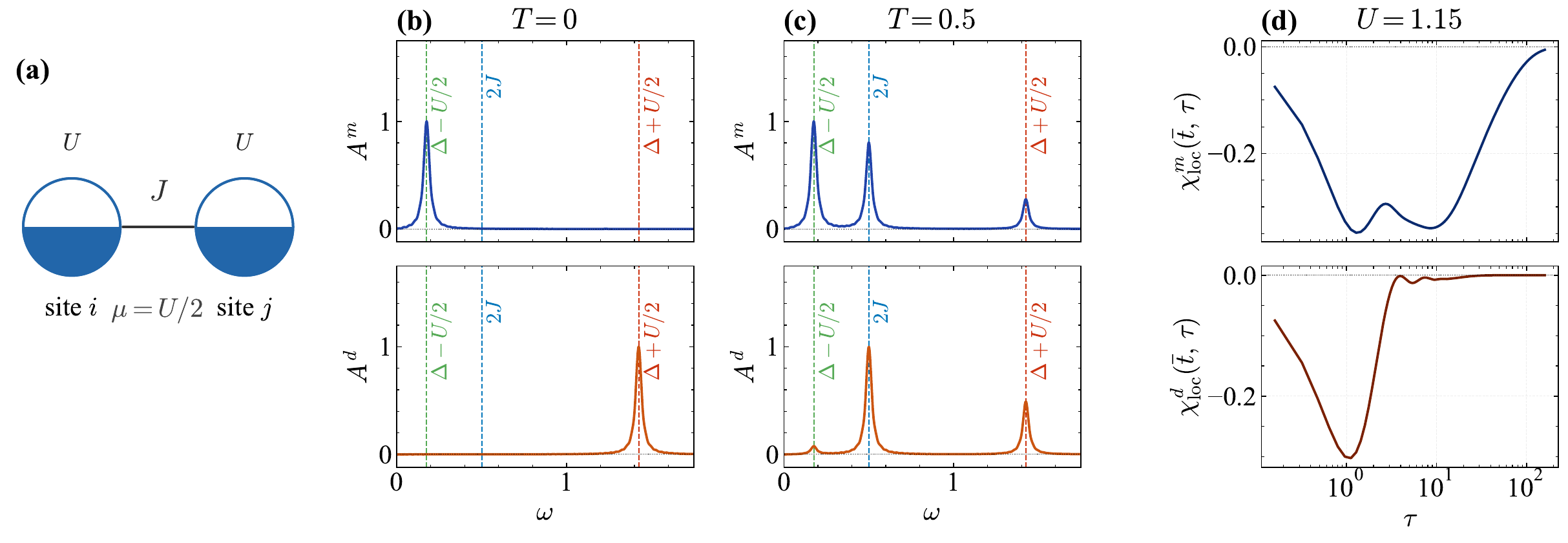}
\caption{(a) Dimer model with onsite interaction $U=1.25$ at half filling, used to identify the microscopic origin of the $\chi^{m}$ and $\chi^{d}$ peaks. (b,c) Dimer spectral functions $A^{m}(\omega)\equiv-\mathrm{Im}\,\chi^{m}(\omega)$ (top) and $A^{d}(\omega)\equiv-\mathrm{Im}\,\chi^{d}(\omega)$ (bottom), normalized to their peak value, for (b) $T=0$ and (c) $T=0.5$. Dashed lines mark $\Delta-U/2$ (green), $2J$ (blue), and $\Delta+U/2$ (red). (d) $\chi^{m}_{\mathrm{loc}}(\bar t,\tau)$ (top) and $\chi^{d}_{\mathrm{loc}}(\bar t,\tau)$ (bottom) from the pulse-driven lattice dynamics at $U=1.15$, at the probe time $\bar t=80$, on a logarithmic $\tau$ axis, for comparison. All quantities in half-bandwidth ($D=4J$) units.}
\label{fig:dimer-toy}
\end{figure*}

To identify the microscopic origin of the two characteristic features observed in the real-time local spin susceptibility $\chi^m_{\rm loc}(\bar t,\tau)$ discussed in Sec.\ref{sec:llm-criterion}, we analyze the simplest interacting system capable of capturing all of them: an isolated symmetric Hubbard dimer at half filling, illustrated in Fig.\ref{fig:dimer-toy}(a). The dimer, with hopping $J$ and onsite interaction $U$, is solved in the particle--hole symmetric case ($\mu=U/2$) using the real-time exact-diagonalization solver \texttt{triqs-realevol}\cite{realevol}, implemented within the \texttt{TRIQS} library\cite{TRIQS2015}. Throughout this appendix we measure energies in the same units of the half-bandwidth (${J=0.25}$) and fix ${U=1.25}$. The resulting eigenstates and eigenenergies, organized by particle number $N$, total spin $S$, and inversion (site-exchange) parity $P$, are listed in Table~\ref{tab:dimer-spectrum} following Refs.~\cite{Avella2004,carrascal2015hubbard}.

The $N=2$ sector contains the half-filled ground state (the bonding singlet) together with three excited multiplets: the spin triplet, a nonbonding singlet, and the antibonding singlet, which forms the dimer analog of the upper Hubbard sub-band. The $N=1$ and $N=3$ sectors correspond to states with one removed or one added electron, respectively, and each consists of bonding and antibonding one-particle levels separated by $2J$, independent of $U$.

\begin{table}[h]
\centering
\caption{Eigenenergy spectrum of the half-filled Hubbard dimer in the particle-hole symmetric case ($\mu=U/2$). Listed are the many-body eigenstates, labeled by the index $\#$ used in Table~\ref{tab:dimer-transitions}, and classified by particle number $N$, total spin $S$, and inversion (site-exchange) parity $P=\pm1$, together with their grand-potential energies $\Omega=E-\mu N$. Here, $\Delta=\sqrt{(U/2)^2+4J^2}$. The ground state (GS) is the bonding singlet.}
\label{tab:dimer-spectrum}
\begin{tabular}{llcccc}
\hline\hline
$\#$ & Sector & State & $S$ & $P$ & $\Omega$ \\
\hline
$1$  & $N=0$ & vacuum & $0$ & $+1$ & $0$ \\
$2$  & $N=4$ & fully occupied & $0$ & $+1$ & $0$ \\
$3$  & $N=1$ & bonding & $1/2$ & $+1$ & $-U/2-J$ \\
$4$  & $N=1$ & antibonding & $1/2$ & $-1$ & $-U/2+J$ \\
$5$  & $N=3$ & antibonding (hole) & $1/2$ & $-1$ & $-U/2-J$ \\
$6$  & $N=3$ & bonding (hole) & $1/2$ & $+1$ & $-U/2+J$ \\
$7$  & $N=2$ & bonding singlet (GS) & $0$ & $+1$ & $-U/2-\Delta$ \\
$8$  & $N=2$ & triplet ($3$-fold) & $1$ & $-1$ & $-U$ \\
$9$  & $N=2$ & nonbonding singlet & $0$ & $-1$ & $0$ \\
$10$ & $N=2$ & antibonding singlet & $0$ & $+1$ & $-U/2+\Delta$ \\
\hline\hline
\end{tabular}
\end{table}

Throughout this appendix we consider the on-site susceptibilities $\chi^{m}(t)=-i\theta(t)\langle[S_z^{1}(t),S_z^{1}(0)]\rangle$ and $\chi^{d}(t)=-i\theta(t)\langle[n_{1}(t),n_{1}(0)]\rangle$, where $S_z^{i}=(n_{i\uparrow}-n_{i\downarrow})/2$ and $n_{i}=n_{i\uparrow}+n_{i\downarrow}$ are the spin and charge densities on site $i$.
These are the dimer analogs of the local lattice quantities $\chi^{m,d}_{\rm loc}$ of the main text.
Their different excitation spectra follow from the symmetry of the corresponding local operators.
Decomposing them into uniform and staggered parts, ${S_z^{1}=(S_z^{\rm tot}+S_z^{\rm st})/2}$ and ${n_{1}=(n^{\rm tot}+n^{\rm st})/2}$ with ${S_z^{\rm st}=S_z^{1}-S_z^{2}}$, the uniform components commute with the Hamiltonian and, thus, are constant, so that the entire dynamical response is carried by the staggered components. These are odd under site exchange and therefore connect only states of opposite inversion parity, ${\Delta P=-1}$. Additionally, both operators conserve the particle number, $S_z^{\rm st}$ allows ${\Delta S=0,\pm1}$ but vanishes between two singlets, while $n^{\rm st}$ preserves $S$. Together, these rules completely determines the excitation spectra of the spin and charge susceptibilities.

Figure~\ref{fig:dimer-toy}(b) shows the spectral functions of spin and charge excitations, $A^{m}(\omega)\equiv-\mathrm{Im}\,\chi^{m}(\omega)$ and $A^{d}(\omega)\equiv-\mathrm{Im}\,\chi^{d}(\omega)$, for the dimer at $T=0$.
Since only the ground state is thermally populated, the spectra contain only transitions originating from the bonding singlet.
The combined spin and spatial selection rules further restrict these to a single allowed transition in each response channel, highlighted in bold in Table~\ref{tab:dimer-transitions}.
The spin excitation spectrum therefore consists of a single peak corresponding to the bonding-singlet$\rightarrow$triplet transition, with excitation energy
\begin{equation}
\Omega_{\mathrm{triplet}}-\Omega_{\mathrm{GS}}
=\Delta-\frac{U}{2}
\;\xrightarrow{\,U\gg J\,}\;
\frac{4J^2}{U},
\end{equation}
where
\begin{equation}
\Delta \equiv \sqrt{\left(\frac{U}{2}\right)^2 + 4J^2}.
\end{equation}
In contrast, the charge excitation spectrum contains a single peak corresponding to the bonding-singlet$\rightarrow$nonbonding-singlet transition, with excitation energy
\begin{equation}
\Omega_{\mathrm{NB}}-\Omega_{\mathrm{GS}}
=\Delta+\frac{U}{2}
\;\xrightarrow{\,U\gg J\,}\;
U.
\end{equation}

\begin{table}[h]
\centering
\caption{Transitions between the dimer eigenstates listed in Table~\ref{tab:dimer-spectrum}, labeled by their state indices $\#$, together with their excitation energies $\omega$ and the response channel in which they carry spectral weight. The selection rules require $\Delta P=-1$ in both channels. In addition, the spin susceptibility $\chi^m$ allows transitions with $\Delta S=0,\pm1$, except between two singlet states, whereas the charge susceptibility $\chi^d$ requires $\Delta S=0$. The transitions shown in bold are the only ones allowed at $T=0$.}
\label{tab:dimer-transitions}
\begin{tabular}{lcc}
\hline\hline
Transition & Channel & $\omega$ \\
\hline
$\boldsymbol{7\to8}$                        & \textbf{spin}              & $\boldsymbol{\Delta-U/2}$ \\
$9\to10$         & charge          & $\Delta-U/2$ \\
$3\to4$ & spin, charge      & $2J$ \\
$5\to6$ & spin, charge      & $2J$ \\
$8\to10$          & spin              & $\Delta+U/2$ \\
$\boldsymbol{7\to9}$                        & \textbf{charge}            & $\boldsymbol{\Delta+U/2}$ \\
\hline\hline
\end{tabular}
\end{table}

As the temperature is raised to $T=2$, excited states become thermally populated, opening additional transitions. All transitions between the eigenstates of Table~\ref{tab:dimer-spectrum}, together with their excitation energies and the response channel in which they carry spectral weight, are listed in Table~\ref{tab:dimer-transitions}. The resulting spectra are shown in Fig.~\ref{fig:dimer-toy}(c).

The first new feature is a peak at $\omega=2J$, independent of $U$, originating from bonding-to-antibonding single-particle excitations within the thermally populated $N=1$ and $N=3$ sectors. This excitation carries spectral weight in both the spin and charge channels. The second is a peak at $\omega=\Delta+U/2\rightarrow U$ in $A^{m}$, arising from the thermally activated triplet$\rightarrow$antibonding-singlet transition. These two high-energy spin excitations, at energies $2J$ and $\Delta+U/2$, are the dimer counterparts of the high-energy magnetic excitations reported for the cuprate model in Ref.~\cite{Stepanov2019}.

In the charge channel, the elevated temperature likewise opens a transition at the superexchange energy, corresponding to the nonbonding-singlet$\rightarrow$antibonding-singlet excitation. However, since both states lie approximately $U$ above the ground state, their thermal occupation remains exponentially small, and the associated spectral weight is negligible.

The two spectra therefore differ, in practice, only through the feature at $\omega=\Delta-U/2\rightarrow4J^2/U$, which appears in $A^{m}$ already at $T=0$ but is essentially absent from $A^{d}$. 
This excitation corresponds to the Heisenberg superexchange scale $J_{\rm ex}$, and its exclusive appearance in the spin channel provides the spectroscopic signature of local-moment formation. 
This interpretation is consistent with the behavior of the local spin susceptibility discussed in the main text. For $U=1.15$, $\chi^m_{\mathrm{loc}}(\tau)$ exhibits a well-separated long-time peak (Fig.~\ref{fig:dimer-toy}), whose characteristic timescale is set by $J_{\rm ex}^{-1}$. 
We therefore associate this slow component with the dynamics of interacting local moments. By contrast, the excitation at $\omega=2J$ originates from bonding-to-antibonding hopping processes and appears in both the spin and charge channels. 
Its higher excitation energy corresponds to a shorter characteristic timescale, consistent with the fast peak observed in the local susceptibilities and with the rapid relaxation process reported in Ref.~\cite{sayad_relaxation_2016}. 
Finally, the excitation at $\omega\simeq U$ occurs on an even shorter timescale and therefore lies beyond the temporal resolution accessible in our nonequilibrium calculations.


\end{document}